\documentclass[a4paper,11pt]{article}
\usepackage{jinstpub} 
\usepackage{lineno}
\usepackage{upgreek}

\usepackage{xspace}
\newcommand{\TM}{\emph{Topmetal}\xspace}
\newcommand{\TMIIm}{\mbox{\emph{Topmetal-II\raise0.5ex\hbox{-}}}\xspace}
\newcommand{\TMS}{\mbox{\emph{Topmetal-S}}\xspace}

\title{Ton-scale Xenon Gas TPC for $0\nu\beta\beta$ Search at Atmospheric Pressure}

\author[1]{Y.~Mei,}
\author[1]{K.~Mistry,}
\author[1]{D.R.~Nygren,}
\affiliation[1]{
Department of Physics, University of Texas at Arlington, Arlington, TX 76019, USA}
\emailAdd{yuan.mei@uta.edu}
\emailAdd{krishan.mistry@uta.edu}
\emailAdd{nygren@uta.edu}

\abstract{
We explore aspects of an unorthodox ton-scale xenon gas time projection chamber detector operated at normal temperature and pressure (NTP), designed for $0\nu\beta\beta$ discovery at $\sim10^{27}$ year sensitivity.  At fixed active mass, a greater transparency to $\gamma$-ray backgrounds and better track clarity exists at NTP relative to higher density. Operation at NTP also alleviates difficulties with pressure containment and significantly reduces HV requirements. With metre-scale electron tracks at a few MeV, event topology can be used to efficiently reject the copious $\gamma$-ray backgrounds by factors ranging from 550--1200, largely compensating for the greater surface area of the detector. Event topology is captured from anode plane signals combined with sensing the secondary ion component arriving at the cathode plane.  A molecular gas admixture limits diffusion of the secondary electron tracks to less than one cm rms and provides stable proportional avalanche gain for measurement of energy. An energy resolution $\delta$E/E $\leq$ 1\% FWHM seems possible. A 100 kg scale demonstrator for this detector technology could lead directly to a ton-scale search.

}

\begin{document}
\maketitle
\flushbottom

\section{Context: Ton-scale Xenon Experiment}
\label{sec:intro}
Since the possibility of double-beta decay~\cite{goeppertmayer1935double} was first postulated a century ago, the search for $0\nu\beta\beta$~\cite{furry1939transition} in any of several candidate isotopes remains a compelling but unfulfilled scientific goal. A clear observation would confirm the Majorana nature of the neutrino and also add support for a leptogenesis scenario of matter-antimatter asymmetry in the universe~\cite{FUKUGITA198645}. Theoretical considerations suggest that a ton-scale search may achieve success. The experimental challenge is clear: detect and identify true signal events with the highest practical efficiency and clarity while identifying and rejecting background events at a level such that observation of a small number of signal events constitutes a credible discovery. In any search, the capability for demonstration that ``signal follows isotope'' -- through easy interchange of isotopically enriched active mass with depleted / natural active mass -- is, in our view, an essential feature for a robust claim of discovery of such fundamental importance.

A pioneering $0\nu\beta\beta$ search was performed in the Gotthard tunnel in the 1990's~\cite{PhysRevD.48.1009,Luscher:1998sd} with a high-pressure xenon gas time projection chamber (TPC). The TPC held 3.3 kg of $^{136}$Xe with 3.9\% methane at a pressure of 5 bar using a modified multi-wire proportional chamber design to visualize tracks and measure energy. This experiment established a limit $T_{1/2} > 4.4 \times 10^{23}$ years and introduced the use of event topology to distinguish ``head-tail'' differences in track appearance, as the higher energy origin of an electron track is generally much ``straighter'' than the lower energy track end around 100 keV, where multiple scattering is much more prominent. The basic idea is that the two-electron signal events generally have ``blobs'' at both ends of a continuous track, whereas single-electron background tracks generally display only a single blob. It is within these scattering effects that background discrimination power lies; multiple scattering has here become a virtue. Our scenario attempts to exploit this intrinsic event information -- energy and topology -- to their practical discrimination power limits.   Alternative approaches against backgrounds that invoke self-shielding with $^{136}$Xe will require a factor of several times more isotope and will face feasibility challenges due to cost and isotope availability issues. 

The ongoing KamLAND-Zen experiment, with 745 kg of $^{136}$Xe dissolved in liquid scintillator, has produced the best current xenon limits: $T_{1/2} > 3.8 \times 10^{26}$ years~\cite{kamlandzen2025}. The KamLAND-Zen collaboration is modifying the detector with a goal of improving the energy resolution to below 6$\%$ full-width-at-half-maximum (FWHM)~\cite{KamLAND2_Prototype_2025}, thereby reducing background contributions from radioactivity and the two-neutrino decay mode. The scenario we describe here is in no sense intended to suggest that the ongoing KamLAND-Zen experiment may not succeed in observing a faint signal beyond its current limit. The signal, if any, will initially be weak statistically; any background contribution, if present, would weaken significance.   

The widely accepted background index requirement of $<$ 0.1 event / ton-year in the energy region of interest (ROI) is an especially daunting challenge\footnote{For context, one kg of the ``detector'' generates less than one false signal event in 10,000 years.}. For comparison, the expected $0\nu\beta\beta$ event rates in $^{136}$Xe with $10^{27}$ and $10^{28}$ year half-lives are about 3 and 0.3 events/ton-year, respectively. Radiopurity of ton-scale detector systems at this level of background rejection has not yet been demonstrated. In addition to achieving adequate rejection of radioactivity-induced backgrounds, an energy resolution of  $\delta$E/E $\leq$ 1\% FWHM at the $^{136}$Xe $0\nu\beta\beta$ Q-value of 2.458 MeV may be required to eliminate a contribution from $2\nu\beta\beta$ decays, otherwise indistinguishable.

In scaling any xenon experiment toward the ton-scale or beyond, including a TPC approach using xenon gas, rejection of backgrounds is clearly predominant in experimental design. NEXT-100~\cite{next100} is the exemplar of making a xenon gas TPC detector as small as practical, setting nominal operating pressure at 13.5 bars. 

The impacts of varying gas density can be found in Ref.~\cite{OptimalParamsGXeTPC}. Our basic premise in this work is that not only do events contain enough information sufficient to permit a credible search at the ton-scale based on a xenon gas TPC, but the TPC can be successfully operated at normal temperature and pressure (NTP). This choice allows simplified detector construction and shielding with relatively conventional materials. 

In the absence of hard scatterings that produce energetic $\delta$-rays and Bremsstrahlung, the combined track length of the primary electrons can reach $\sim$360 cm at NTP. As shown in the Gotthard TPC experiment, multiple Coulomb scattering of electrons in xenon leads to highly variable, complex, and tangled electron trajectories.   Complicating this picture further is the frequent occurrence of $\delta$-rays, which may occur anywhere along the trajectory. An especially challenging case is the occurrence of an energetic $\delta$-ray near the beginning of a one-electron track, which can then lead to complex two-electron topologies that may closely mimic true signal events. 

We consider operation at NTP, hereafter referred to as the ``NTP-TPC''. Ionization electrons arriving at the modified multi-wire proportional chamber (MWPC) anode plane are amplified by proportional avalanche gain, likely between 300 - 500.  A molecular component such as propylene is added at the few \% level to optimize energy resolution and operational stability, and to reduce electron diffusion during drift. If necessary, an additional molecular component, trimethylamine (TMA) --- which displays a strong Penning effect, could be included to reduce HV requirements substantially.  Conventional admixtures such as xenon plus CO$_2$ are also attractive possibilities but are likely to yield slightly poorer energy resolution (see Section \ref{sec:eres}).

The anode plane provides detailed ($x$, $t$) and energy information while secondary ions arriving at the cathode plane provide associated ($y$, $t$) information. Even with possible ambiguities due to spatial projections, the contiguous and mainly isolated track elements allow complete event reconstruction in 3-D. Placement in $z$ is determined by the diffusion of the track image during drift.

Although our NTP  scenario is quite unorthodox, it appears to offer important advantages relative to higher densities, such as more manageable high-voltage and gas containment systems, but also more scope for eliminating backgrounds with advanced reconstruction tools enabled by the more detailed track topologies. Several critical R\&D questions would need to be fully addressed to realize this detector, including (i) whether the desired energy resolution $\delta$E/E $\leq$ 1\% FWHM can be achieved, (ii) whether a robust 3-D event reconstruction can be made without the use of the primary scintillation signal (which is quenched with the introduction of molecular gas additives), (iii) whether event topology can provide the discrimination power necessary, and (iv) whether an affordable, practical, and background-free experiment is possible without exotic or costly materials.

We take note that more than one ton of the isotope $^{136}$Xe already exists worldwide in various locations~\cite{kamlandzen2025,PhysRevLett.123.161802,next100}, reinforcing motivation for a possibly definitive ton-scale experimental search in the relatively near term. Here we present an explicit design approach for a ton-scale xenon gas TPC system at NTP. Our results thus far suggest that an opportunity could exist at the ton- or plausibly even at a few-ton scale. Further R\&D appears well motivated.

The paper is structured as follows. Section~\ref{sec:gasdense} discusses the impacts and benefits of operating at NTP. Section~\ref{sec:tpc} covers the aspects of the TPC design, including the geometry, gas system, field cage, and anode and cathode plane design. Section~\ref{sec:reco} discusses the use of electron and ion signals for 3-D event reconstruction. Section~\ref{sec:eres} presents the scheme for achieving the required energy resolution of less than 1{\% FWHM}. Section~\ref{sec:topology} investigates the background rejection power attainable through topology, using both algorithmic and machine learning (ML) techniques. Section~\ref{sec:demonstrator} details a potential demonstrator experiment. Finally, Section~\ref{sec:summary} offers a summary with future prospects.


\section{Gas Density: Impacts of Choice}\label{sec:gasdense}
The choice of NTP seems in obvious conflict with common sense, which would presume that the detector should be made as small as possible to minimize surface area.  However, the interplay of the many variables such as event clarity, intrinsic background due to detector size, the signal/background containment and practical considerations reveals a more promising outlook, apparently competitive with current approaches. 

With constant detector geometry and fixed active mass, reducing gas density from, for example,  15 bar to NTP increases linear dimensions by a factor of $15^{1/3} = 2.47$. Were all other things equal, $\gamma$-ray backgrounds would seem to increase by area, a factor of $15^{2/3}\sim6.1$. However, for fixed mass and detector geometry, the conversion of MeV $\gamma$-rays in gas at NTP is much smaller, since the density is reduced by a factor of 15 while the absorption length is only a factor of 2.47 larger. Since the event ``size'' is proportional to 1/$\rho$ while the detector dimensions scale as $\rho^{-1/3}$, a substantially larger fraction of events will penetrate the fiducial boundaries at NTP, an adverse effect leading to lower overall containment efficiency affecting both signal and backgrounds. 

Track ``clarity''  is key to effective discrimination between signal and background events, and may be characterized roughly by a comparison of the total track length, $l$, (stretched out)  with the maximum diffusion of the secondary electrons during drift to the anode. For a given primary electron energy, in the absence of radiative scattering, the total track length $l$ is inversely proportional to density, 1/$\rho$.  The diffusion after drift time $t$ is given by $\sigma^2 = 2tD = 2tvd/3$, where $D$, the diffusion coefficient, is the product of $v$, electron speed (given by agitation energy), and $d$, the mean distance between electron-atom scatters in the gas. $d$ is also inversely proportional to density, 1/$\rho$.  The maximum drift time $T$ is given by $L/v_{\textnormal{drift}}$, where $L$ is some measure of the detector geometry and $v_{\textnormal{drift}}$ is the electron drift velocity. For fixed target mass, $L$ is inversely proportional to the cube root of density, $\rho^{-1/3}$. Thus $\sigma_{\textnormal{max}}^2 $ is proportional to $\rho^{-4/3}$  and  a dimensionless figure of merit, FOM =  $l^2  /\sigma_{\textnormal{max}}^2 $ is  proportional to  $\rho^{-2/3}$. In this example, relative to 15 bars, the FOM is better at NTP by a factor of 6.1. A significant benefit at low density is apparent, lending support for operation at NTP.

At NTP, the drift field is reduced by a factor of 15 while the maximum drift length $L$ is increased only by $15^{1/3}$ = 2.47.  The high voltage (HV) requirement, compared with that at 15 bars, is thus reduced by a factor of $15^{2/3}$ = 6.1, leading to a requirement of less than 20 kV at NTP.  Such low operational voltages are more compatible with conventional high voltage supplies, insulation, and corona prevention techniques. 

Furthermore, attachment by electronegative impurities is greatly reduced at NTP, as that process is promoted by three-body collisions needed to satisfy energy-momentum balance, leading to a $\rho^2$ dependence. Reduced by a factor of 1/225, attachment at NTP is likely negligible. Due to the anticipated very low background ionization rate, the ionic space charge that might distort the drift field in the TPC volume has an insignificant topological impact. Volume recombination is negligible.  At NTP, K-shell fluorescence x-rays from photoproduction by background gammas are more separated spatially and may be experimentally more recoverable as tags of the process.  Heavy particle tracks and $\alpha$-particles will be topologically completely distinct, inducing a negligible background.

Initial explorations of these dependencies with gas density were carried out with an analysis utilizing a simple algorithmic approach in Ref.~\cite{OptimalParamsGXeTPC}, where it appeared that NTP operation had approximately a factor of four higher background compared with higher pressures. This reduced performance was ultimately due to the lower containment efficiency limiting the strength at which cuts could be applied; however, it was clear that this analysis did not effectively utilize the full topological information available. This aspect is revisited in this work at NTP with more advanced reconstruction tools that better capture information within events.

A robust method of placement of an event in 3-D space is essential to avoid backgrounds arising from anode, cathode, or field cage surfaces. In practice, the $x$, $y$, and $z$ placement schemes must be able to ensure a distance of $\sim$10 cm or more exists from any solid surface to any part of a reconstructed event track to suitably eliminate radon backgrounds and calibrate the detector. 

A design at NTP permits gas containment, uniquely, with a mechanically thin enclosure fabricated in situ with manageably small pieces.  Mechanical containment and management of gas at NTP is not only much simpler and much cheaper, but also far less risky since rupture with sudden massive loss can be made implausible. As a result, we suggest that an attractive pathway for a ton-scale detector to be realized in a timely manner, with manageable difficulty, and at reasonable cost might be an implementation at NTP. 


\section{Experimental Conception}\label{sec:tpc}

\subsection{TPC Design: Symmetric and Inverted Geometry}

\begin{figure}[hbt]
\centering
\includegraphics[width=0.85\textwidth]{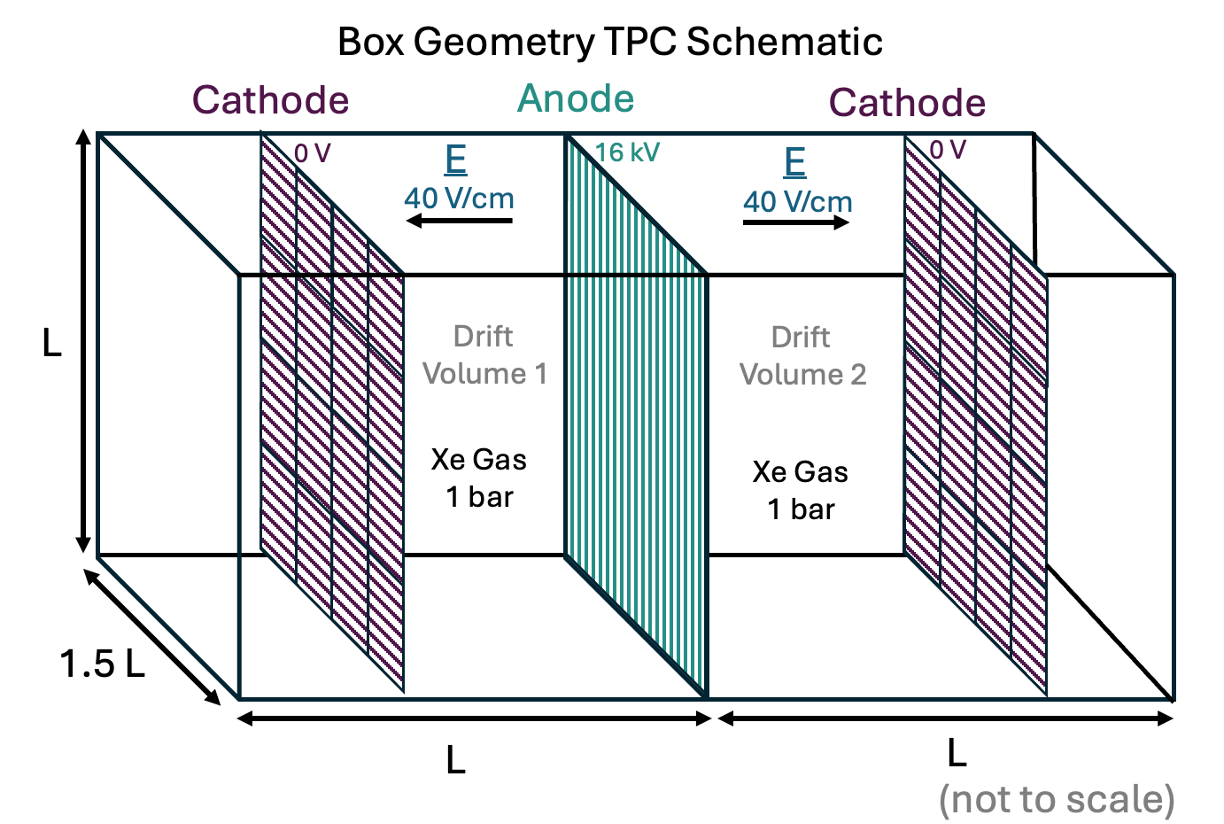}
\caption{\label{fig:tpc} A diagram of the box geometry for the TPC consisting of a central anode reading out ($x$, $t$) information and two cathode planes reading out ($y$, $t$) information. }
\end{figure}

Classically, a symmetric TPC incorporates two anode planes facing a common cathode plane at the center of symmetry. The common cathode plane is sufficiently materially thin that events passing between halves are negligibly compromised, thereby preserving fiducial volume and event quality. Here, the NTP-TPC places a common, thin anode plane at the central point with cathode planes at the opposite surfaces, a still symmetric but ``inverted'' TPC geometry similar to the NEXT-CRAB geometry~\cite{CRAB0} (see Fig.~\ref{fig:tpc}).  This unorthodox approach introduces some real complications, discussed further below, but appears to provide a more practical solution for realizing the desired energy resolution of $\delta$E/E $\leq$ 1\% FWHM at the $^{136}$Xe $0\nu\beta\beta$ Q-value. 

\begin{figure}[hbt]
\centering
\includegraphics[width=0.85\textwidth]{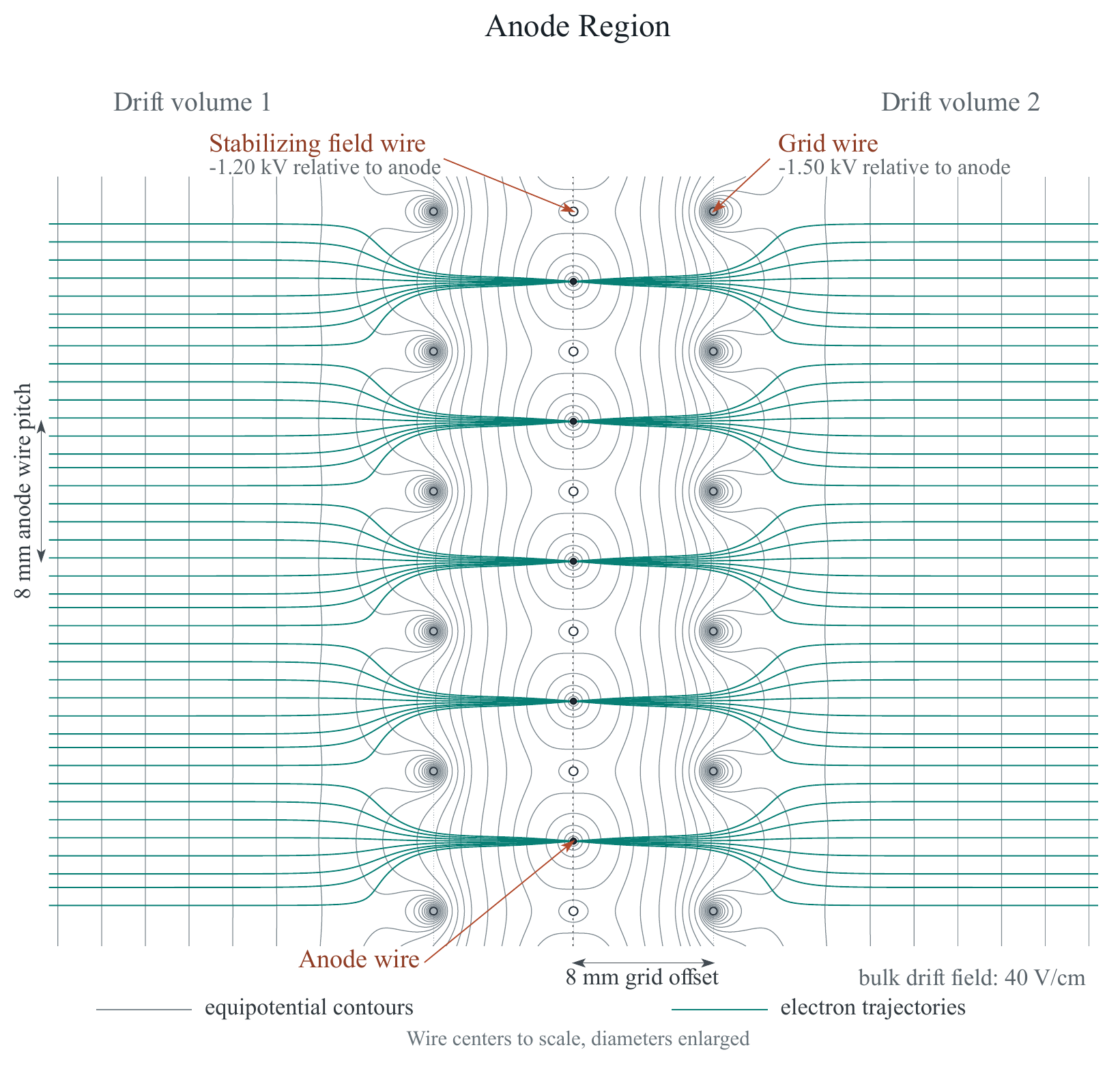}
\caption{\label{fig:AnodePlane} The anode plane wire configuration, showing symmetry between drift volumes. The field wires stabilize against even-odd displacement and circularize the electric field around the anode wire. The anode wire spacing is expected to be about 8~mm. The spacing from anode wires to grid wires is likely also 8~mm. }
\end{figure}

The anode plane is a single plane of vertically hanging wires under tension (See Fig.~\ref{fig:AnodePlane}), reminiscent of the Gotthard TPC and the PEP-4 TPC geometries~\cite{Luscher:1998sd,PEP4}, but now in a symmetric configuration with two planes of grid wires. Geometric symmetry guarantees that electrons arriving from either half of the TPC are amplified equally and also provides greater resilience against electrostatic instabilities. Each anode wire provides a stream of ($x$, $t$) signals. Proportional avalanche gain, despite unavoidable fluctuations that limit energy resolution, appears to provide acceptable performance, as discussed below in Section~\ref{sec:eres}.

In addition to electrical signals for energy measurement at the anode plane, the simultaneously generated ionic components from the avalanche migrate to the cathode planes, providing ($y$, $t$) information. A modest spatial resolution penalty is incurred due to the additional diffusion of the ionic component during transit to the cathode plane. In a classic TPC design, the ($y$, $t$) information would be captured from proximate cathode pad arrays, but here we choose to prioritize energy resolution performance and use a symmetric anode plane in inverted TPC geometry.

A gate closed after the electron trigger could shorten the ion-induced signal, but it would also intercept the secondary ions used here for the remote-cathode $y$ measurement. Reading $y$ from proximate pads or strips is a viable alternative, but it requires a different anode-plane architecture with additional structures and extremely precise geometric tolerances near the gain region; it is therefore not the baseline considered here.

Event topology capture is thus enabled using a combination of the anode ($x$, $t$) and cathode ($y$, $t$) information.  Energy measurement and topology capture functions are both derived from avalanche gain but sensed with oppositely charged carriers.

In pure xenon, a sub-$\upmu$s primary scintillation signal, $S_1$, is available to identify the instant time of origin, $t_0$, of an energetic event. Measurement of $S_1$ permits a highly accurate placement of an event in $z$ within the active volume.  However, detection of an $S_1$ signal is not feasible here because the molecular additives quench $S_1$ strongly, and, even if quenching were minimal, the required area of radiopure photodetection would be impractically large. Without an explicit $t_0$, we must rely on other methods that can provide an equivalent estimator. 

The cathode planes are attached to detector walls opposite the anode plane and operated at ground potential for safety. The baseline design for the cathode planes is a matrix of sub-planes equipped with cathode strips oriented horizontally (see Section~\ref{subsec:cathode}). These elements provide a stream of ($y$, $t$) signals that enable 3-D reconstruction of event topology (see Section~\ref{sec:reco}). In this NTP-TPC design concept, most electronics associated with signal capture can be placed external to the detector.

For the $x$ coordinate, the continuity of anode wires up to the edge of the active volume guarantees visibility up to the field cage edge.  The field cage potential gradient can be chosen to be slightly non-linear to ensure that any ionization emerging from the field cage surface will be drawn to an active anode wire. 


\subsection{Detector Shape and Size}

We assume the availability of a fixed target mass of one metric ton of $^{136}$Xe. An obvious geometry is a simple octagonal cylinder, forming a symmetric NTP-TPC of diameter $L$, total length $2L$ (convenient but not necessarily optimum).  $L$ is approximately equal to the maximum drift length. The density of xenon gas at NTP is $\rho$ = 5.487 kg/m$^3$, which leads to a volume $V$ = 182 m$^3$ for 1000 kg mass.  The approximate dimensions for an octagonal cylinder are given by $1000 = M \approx \rho \times \pi \times (L/2)^2 \times L \times 2$, yielding L = 4.9 m. 

Alternatively, we may consider a simple box-style symmetric TPC with vertical dimension $y = L$, width $x = 1.5L$, and in drift direction $z = L$, yielding overall length $2L$ (see Fig.~\ref{fig:tpc}). For ton-scale at NTP, $L = 3.93$~m. In either box or cylinder shape, the NTP-TPC has a huge surface area, about 200 m$^2$.  This area must be covered with a thick layer of sufficiently radiopure absorber for external $\gamma$-rays. Copper is the obvious candidate for this shielding. The copper area needed for the one bar scenario is approximately a factor of $15^{2/3}$ = 6 more than for a 15 bar version\footnote{Perhaps the huge quantity of copper needed could be “borrowed” from a benevolent entity formed to provide it and which recovers costs of investment later, as done for the xenon in the LZ experiment~\cite{hickok2022slac}. Current costs of copper are $\sim$13k USD/ton~\cite{lme2026copper}, resulting in a total cost in the range of 2--3M USD for 10-12 cm thickness copper shielding. }. In addition, a substantial layer of hydrogenated material is needed to suppress entry of neutrons into the active volume, where they could produce $^{137}$Xe. In the absence of a detailed design and site, we must defer study of the extensive radiopurity requirements.

Although the easier box-style NTP-TPC will have noticeably smaller geometric containment efficiency than a quasi-cylinder, the attractive simplicity could make it the preferable design choice; we choose the box style for further discussion. Should more $^{136}$Xe become available, simply increasing the dimensions of $x$ in the box geometry by a suitable factor at the design stage provides a scalable solution. For any scale factor in $x$,  we choose an $E/P$ of 40~V/cm/bar to minimize diffusion. Only $\sim$400 cm $\times$ 40 V/cm = 16 kV is needed at NTP for the box-style NTP-TPC. For ton-scale cylindrical geometry and a pressure of 15 bar, the TPC would require $\sim$100 kV. 


\subsection{Gas Containment and Related Systems}
The NTP-TPC containment vessel can be constructed in place, using radiopure, relatively lightweight pieces such as aluminum, aluminized PMMA panels, or perhaps a graphene composite. No large, extremely massive pieces need to be transported to the actual underground site by elevator and through long tunnels. A "double-hull" design would be natural, adding strength and enabling reliable leak detection by circulating argon in the interstices. 

The gas system may utilize a setup similar to the ALICE Transition Radiation Detector (TRD), which employed a MWPC with Xe/CO$_2$ (85-15 ratio) at atmospheric pressure~\cite{ALICETRD}. Commissioning can be done with a flush of pure argon or CO$_2$  until oxygen and water levels are asymptotically low and detector systems are tested. For operation with xenon, this fill is then replaced by a helium blanket introduced smoothly from above. Finally, a transition to the enriched xenon + molecular admixture is done by carefully pushing the helium up and out.  

Xenon recovery is accomplished by reversing the process; the presence of helium in the xenon mix is benign and easily separated cryogenically. This process can also be used to remove the slow buildup of nitrogen in the gas. The subsystems for helium and xenon storage, distribution during fill, and recovery must be highly integrated for simultaneous operation, with particular care given to avoiding any significant over- or under-pressure. If engineering considerations indicate that a catastrophic breach of gas containment is sufficiently improbable that a crash recovery system is not needed, a considerable saving in complexity and cost may be possible. 

Purification of xenon is mainly concerned with the removal of water and oxygen, as nitrogen and other contaminants are of secondary importance. Here, with molecular additives, a recirculating system for continuous purification will be more complicated if hot getters are needed. In that case, the molecular additives will almost certainly need to be removed either cryogenically or via capillary polyimide tube membranes~\cite{ALICETRD} from the xenon prior to exposure to hot getters. 


\subsection{Field Cages}
Since the HV requirement is a modest 16 kV, each 0.5 cm step in the field cage supports potential differences of only 20~V. With these voltage requirements, the field cages could be made on a thin plastic foil substrate with continuous double-sided copper traces, stepped and wide enough to overlap slightly to prevent electric field leak-through. This can be manufactured as a roll, where the field cage foils would be applied directly to the TPC structure if made of plastic such as poly(methyl methacrylate) (PMMA). If the TPC is a metallic structure, thin layers of polyethylene with adequate overlap could provide the necessary HV insulation. 


\subsection{Anode Plane: One Symmetric MWPC-Style Array}\label{subsec:anode}
As depicted in Fig.~\ref{fig:AnodePlane}, the anode plane is a curtain of hanging wires under suitable tension to maintain electro-mechanical stability.  An anode wire spacing of 8 mm is chosen to be comparable to the diffusion of electrons (about 3 mm/$\sqrt{\text{drift distance [m]}}$) after drifting several meters in xenon + molecular admixture at NTP. The spatial quantization imposed by the discrete 8 mm spacing of wires seems unlikely to compromise the topological information since a 2.5 MeV track length will be sampled on the order of 200--300 times.  A wider spacing such as 10 mm, if necessary, seems benign.

As the wires are several meters long, additional “field” wires at an appropriate potential are needed to circularize equipotentials near the anode wires, to stabilize the entire anode array against even-odd electrostatic instabilities\footnote{Drift chambers such as the CDF central tracking chamber (CTC) had wires over 3~m in length with similar gain and spacing, and higher fields in a less supportive geometry~\cite{Bedeschi1988}. }. Circularity of equipotentials near the anode wire is also desired for optimum energy resolution, but perhaps some non-circularity could be beneficial for electrostatic stability without compromise to energy resolution.   The anode plane, less than 20 mm thick, is entirely active except for the negligibly small amount of the wire material. 

Two planes of ``grid'' wires are needed to separate the gain region from the two drift volumes.  The proposed anode plane wire geometry with field wires is similar to that of the PEP-4 TPC~\cite{PEP4}, which yielded energy resolution for $^{55}$Fe x-rays nearly equal to that of a cylindrical proportional counter. As the energy resolution performance sought here is so much more stringent, great care must be taken to ensure mechanical precision in all elements of anode plane geometry such as spacing, uniformity of wires, etc. The Gotthard TPC designers made note of the difficulties associated with realizing adequate cathode plane flatness in close proximity to the anode wire plane, to which they attributed a degradation in energy resolution (see Appendix~\ref{appdx:gotthard}). In our ``inverted TPC'' design, electrostatic uniformity is more closely achieved through geometric symmetry and precision naturally obtained by the hanging wires under tension.

The anode plane is operated at +HV to facilitate the operation of the cathode plane at ground, for safety. A single multiconductor HV-capable cable brings in the several low voltages needed to operate high-quality electronic preamps for each wire, all floating near +16 kV. The number of anode wires is $\sim$750 for the box geometry, a bit less for the octagon. 

For energetic electron tracks in xenon at NTP, each channel waveform will, on average, include signals from typically 300 electrons arriving in a few $\upmu$s for 8 mm spacing. For 2.5 MeV electrons, the total track length can be as much as 360 cm. This yields about 200 -- 400 mixed time-space samples, highly dependent on topology, $\delta$-rays, Bremsstrahlung, etc.  With a moderate avalanche gain of $\sim$500, signal size for each channel sample is $\sim$150,000 electrons (24~fC). Electronic noise is expected not to exceed 1000 electrons rms, indicating an S/N ratio in excess of 100. 

Minimally, each preamp drives an optical channel to bring an analog signal out to external processing electronics. Alternatively, the preamp waveform could be digitized internally at $\geq$12 bits to produce a digital bit stream. The digitization rate would be $\sim$1 MHz, corresponding to a sample interval of $\sim$1 mm in $z$ for an expected electron drift velocity of about 1 mm/$\upmu$s. With 8-fold multiplexing, a comfortable optical transmission rate of 100 Mbits/s per fiber with a total of $\sim$100 fibers leads to a manageable raw data rate of $\sim$10 Gbit/s. 

A robust digital signal processing electronic array external to the detector allows full waveform recovery with sophisticated signal processing to provide optimal signal-to-noise (S/N). Almost all data is baseline, essential for maintaining energy measurement stability. Power dissipation within the xenon gas volume could range from less than 4 W for simple buffered electrical-to-optical analog signal transmission to likely less than 40 W for fully digitized and multiplexed optical transmission. As heat is dissipated over several meters, internal cooling may be unnecessary. 


\subsection{Cathode Planes: Secondary-Ion Readout and Primary-Ion R\&D}\label{subsec:cathode}

The cathode plane will receive both the primary ions created along the original track and the avalanche-produced secondary ions that return from the anode.  The primary-ion arrival time would, in principle, provide an independent estimator of the event position in $z$, but the expected charge per topology sample is tiny.  The secondary-ion signal is much larger and provides the $y$ coordinate, but its timing reflects production at the anode and therefore does not recover $t_0$.  For the baseline detector, only the secondary-ion signal is required: it provides the $y$ coordinate, while the event position in $z$ is obtained from electron diffusion as described in Section~\ref{sec:reco}.  Direct primary-ion sensing would provide an additional and independent $z$ estimator, but it is not assumed in the baseline detector concept or in the performance estimates presented here.  The baseline secondary-ion readout and some plausible future primary-ion R\&D options are described below.

\begin{figure}[hbt]
\centering
\includegraphics[width=0.85\textwidth]{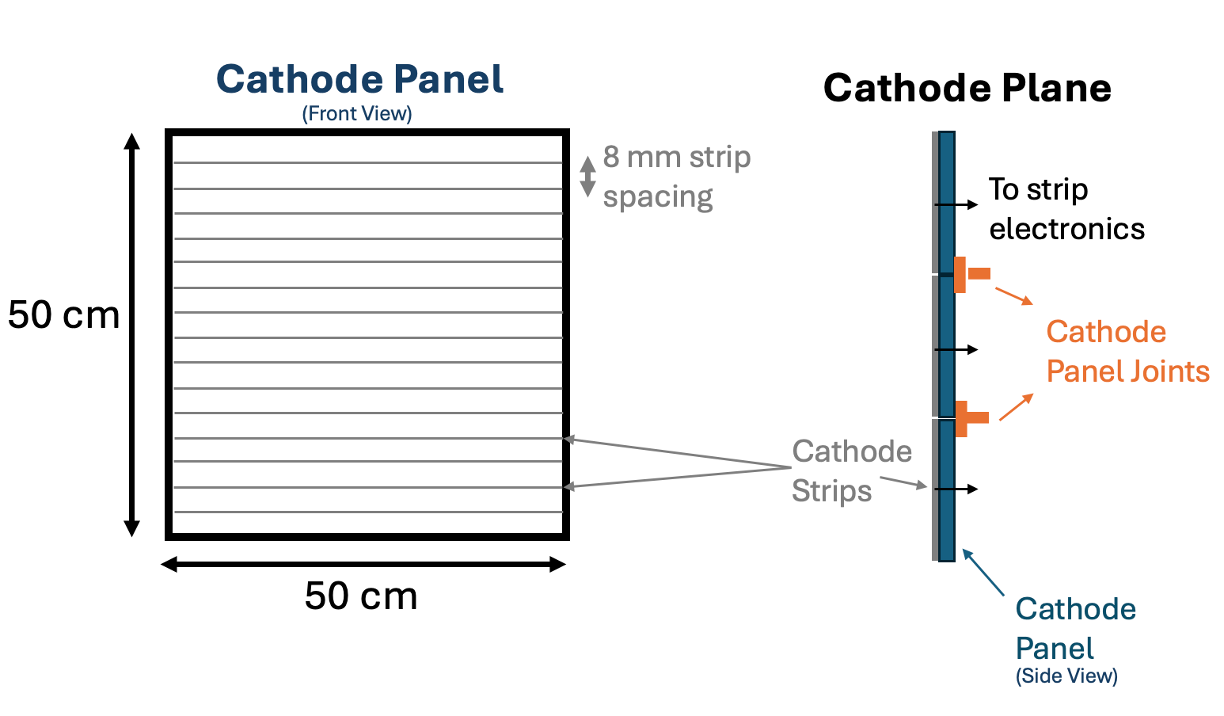}
\caption{\label{fig:cathode} A diagram of (left) a single cathode panel, and (right) how the cathode panels are joined to form the cathode plane.  The strip/panel geometry with exterior CSAs constitutes the baseline secondary-ion readout; direct primary-ion sensing is a separate R\&D objective.}
\end{figure}

For secondary ion readout, the cathode plane is divided into an array of individual ``panels'', which enables the best possible electronic performance as well as facilitating mechanics, construction, testing, and assembly, and ultimately optimizing event reconstruction.  The default readout scheme is shown in Fig.~\ref{fig:cathode}.  Each approximately 8~mm-pitch collector strip spans a 50~cm $\times$ 50~cm panel and connects through a sealed via to an exterior charge-sensitive amplifier (CSA). There are about $\sim$100 panels that are semi-permanently but well-sealed into a large mechanical support grid that comprises a complete cathode plane.

Each panel supports $\sim$60 channels of low-noise electronics optimized for the slow ionic mobility.  Each cathode plane thus supports about 6,000 channels, with all electronics placed exterior to the panel. While the cost of subdivision is a substantially larger channel count, this approach reduces ambiguities in $x$-$y$ matching, facilitates mechanics, and may be necessary to achieve the required electronic S/N.

As a further protection against energetic electron tracks entering the active volume through the cathode panels undetected, the panels could be made of scintillating plastic.  Low-radioactivity photomultiplier tubes (PMTs) placed centrally at each panel would see enough scintillation light to tag charged particles originating from outside the detector.  An alternative photon-detection scheme might use embedded wavelength-shifting fibers.

The conventional direct-charge readout of the plane is intended to measure the avalanche-produced secondary ions.  For $N_{\mathrm{p}}\simeq300$ primary ions per sample, avalanche gain $G=500$, and ion-backflow fraction $f_{\mathrm{IBF}}=0.5$, the returning charge is $Gf_{\mathrm{IBF}}N_{\mathrm{p}}\simeq7.5\times10^{4}$ ions, or approximately 12~fC before sharing.  This avalanche-amplified signal appears compatible with the present strip/panel geometry and a conventional direct-charge CSA architecture, subject to experimental validation of the long-time-constant performance discussed below.  No input-node modulation or other signal-up-conversion mechanism is required for this baseline readout.

Electron and ion readout require different CSAs.  An anode electron packet arrives in a few $\upmu$s; preserving 1~MHz longitudinal sampling calls for effective shaping of order 1--10~$\upmu$s.  Positive ions have a thermal arrival width of approximately 18--30~ms, with realistic effects motivating tests up to 100~ms.  Ion shaping is therefore three to five orders of magnitude slower.  This is not merely a downstream-filter change: the CSA feedback decay $\tau_{\mathrm{f}}=R_{\mathrm{f}}C_{\mathrm{f}}$ must greatly exceed the arrival window.  For an uncorrected constant-current pulse, $\tau_{\mathrm{f}}\gtrsim10\tau_{\mathrm{s}}$ limits ballistic deficit to about 5\%, requiring roughly 0.2--1~s here; percent-level raw retention requires approximately 0.9--5~s.  These unusually long time constants require a dedicated cathode-CSA design and experimental validation, but do not require a fundamentally new signal-transduction principle for the secondary-ion signal. At these long times, Metal-Oxide-Semiconductor Field-Effect Transistor (MOSFET) $1/f$ noise, random-telegraph noise, leakage, feedback-current noise, and $R_{\mathrm{f}}C_{\mathrm{f}}$ stability must be controlled for the secondary-ion readout, but the avalanche-amplified signal provides substantial noise margin.

For direct primary-ion sensing, the signal is only 300~ions, or 0.048~fC, so an unshared signal-to-noise target $R$ corresponds to a system Equivalent Noise Charge (ENC) ceiling of $300/R$~electrons; $R=5$ gives 60~electrons only as an illustrative benchmark.  With the expected capacitance of a full-size collector and an 18--100~ms signal-estimation interval, this primary-ion signal is too weak to be assumed as a realizable direct measurement with a conventional strip--CSA front end.  The required $R$ must ultimately be set by reconstructed hit efficiency and the event-level false-hit allowance.  Further R\&D is therefore required for primary-ion sensing.  The primary-ion R\&D path retains the strip plane but translates its signal above the low-frequency noise before the first transistor, using input-node chopping/commutation, balanced modulation of the approaching ions, or micro-electromechanical systems (MEMS) capture followed by capacitance or nonlinear-resonator frequency transduction.  Unmodulated direct strip--CSA primary-ion sensing remains a useful prototype test, not a detector assumption.  In particular, the baseline $y$ reconstruction and the performance estimates presented in this work do not depend on successful primary-ion sensing.

A monolithic \TM-style~\cite{Mei2020TopmetalPlane} cathode, on which ions land directly on exposed Complementary Metal-Oxide-Semiconductor (CMOS) electrodes, remains a separate future R\&D option that would replace rather than instrument the strip plane.  It is not part of the baseline cathode readout.  \TMIIm measured 13.9-electron ENC at approximately 23~fF input capacitance, 1~ms-FWHM digital shaping, and $\tau_{\mathrm{f}}=7.6$~ms~\cite{An2016TopmetalII}; the \TMS concept simulated a below-30-electron goal for an approximately 5~pF electrode at about 180~$\upmu$s shaping~\cite{Mei2020TopmetalPlane}.  Neither is an ENC result for the default strip load at 18--100~ms.  Appendix~\ref{appdx:ion-sensing} gives the condition-tagged benchmarks, modulation analysis, and quantitative short-shaping energy-resolution calculation.


\section{Event Reconstruction}\label{sec:reco}

The ($x,t_e$) coordinates of an event are directly derived from the anode plane signals. Coordinates  ($y,t_i$) are derived from secondary ion signals at the cathode plane. A synthetic $t_0$ to provide the $z$ coordinate is derived from a global fit to diffusion. The reconstruction of the full 3-D event from 2-D projections in a hybrid strip-wire-plane-readout TPC is not completely trivial due to the inherent ambiguity of matching multiple 2-D wire/strip-plane views into a unique 3-D image. However, this challenge has been successfully addressed by mature, validated algorithms now standard in production for experiments including MicroBooNE and DUNE, for example Pandora and Wire-Cell~\cite{Qian_2018, Acciarri:2018Pandora}. While track topologies are complex and sometimes tangled, unresolvable spatial ambiguities in reconstruction are rare. 


\subsection{Y Coordinate: from Secondary Ion Signals}

As noted in Section~\ref{subsec:cathode}, a sizable fraction $f_{\mathrm{IBF}}=0.5$ of the secondary ions produced in the avalanche follows the drift field lines into the drift volume and travels to the cathode plane.  Each corresponding secondary ion sample is about 75,000 ions, one-half of the amplified electron signal, and is degraded by two stages of diffusion. Nonetheless, since all ions in a sample contribute to an estimate of position, precision is anticipated to be a fraction of the overall diffusion width. 

Elementary calculations indicate a negligible effect for possible space charge effects that might ``puff out'' and degrade the ion images of the tracks as they emerge from the anode plane and drift toward the cathode plane. We note that field-enhanced emission of electrons at the cathode surface per ion must be much less than 1/$N_\textnormal{{si}}$. This requirement should be easily met in the NTP-TPC since the gains in Geiger-Müller counters are very much higher, and must also meet this condition.  

Assuming the absence of complicating ion charge exchange dynamics, the secondary echo allows precise measurement of ion drift velocity. The original $\upmu$s-level timing of the electron signals is still present but is washed out by diffusion of ions over the several-second drift time. 

The secondary ion signal ($y,t_i$) data are used in concert with the anode ($x,t_e$) data to reconstruct the event in 3-D, but without $t_0$. Reconstruction efficiency matching ($x,t$) and ($y,t$) should be robust since the basic topology is a string of contiguous hits with only occasional complexities such as $\delta$-rays, Bremsstrahlung hits, and crossover linkages. Amplitude information is also available to correlate $x$ and $y$ samples. 

We assume that all secondary ions reaching the cathode planes are the single impurity type with the lowest ionization potential, and that charge exchange occurred sufficiently rapidly that the ions maintain an effectively single-valued drift speed.  Ionic clustering could introduce unwanted smearing, but seems unlikely.\footnote{Clustering of xenon ions in pure xenon has been studied as a function of density \cite{PhysRevA.97.062509} indicating that the timing of an ion image might be blurred due to an ill-defined group drift velocity.} If more than one molecular impurity is present with very similar ionization potentials but different mobilities, charge exchange processes will continuously occur and could lead to a smeared-out arrival time at the cathode. 


\subsection{Z Coordinate from Electron Diffusion}
Placement of events in the drift direction $z$ may also be made through measurement of diffusion of the primary electrons, varying as the $\sqrt{}$(drift distance). This has been explored with extended tracks in the context of NEXT-White, with promising results in pure xenon at 10 bar using a Convolutional Neural Network (CNN) approach~\cite{NavarroDiff}. 

The performance of this $z$-reconstruction technique may be impacted due to the reduced diffusion with the introduction of molecular gas additives. We investigate this aspect with the use of a Sparse 3-D CNN to predict the barycentre (mean $z$) and $z_\textnormal{max}$ (max $z$ hit) position for events with 5\% CO$_2$ admixture. Details of the network are given in Appendix~\ref{appdx:ML}.

\begin{figure}[hbt]
\centering
\includegraphics[width=\textwidth]{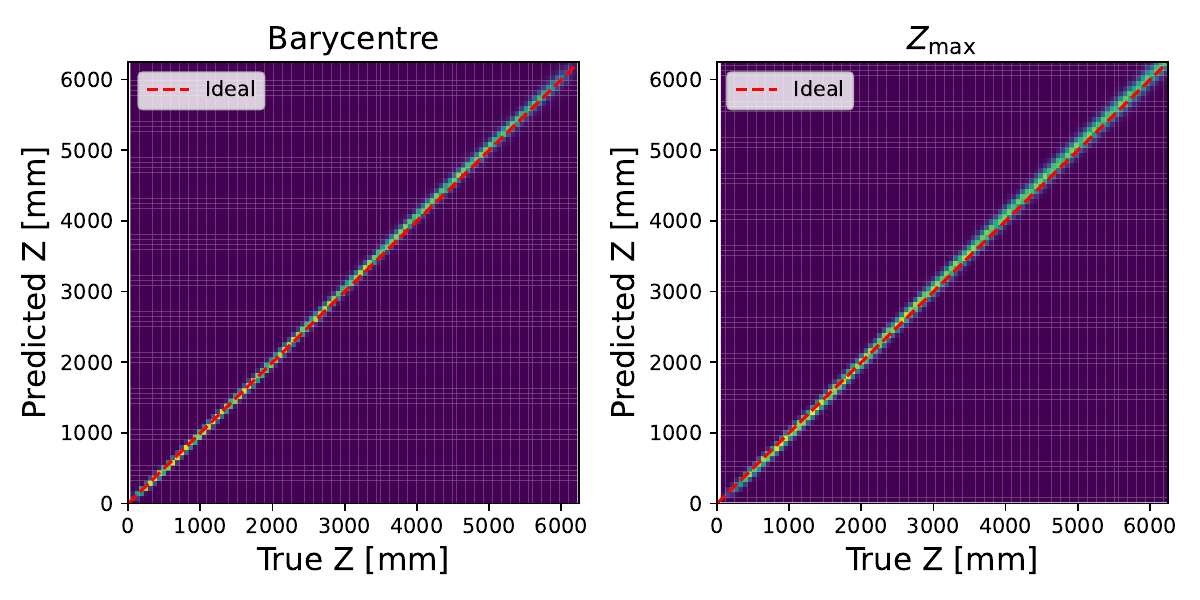}
\caption{\label{fig:CNNZPos} Predictions by the CNN for the barycentre and  $z_\textnormal{max}$ on the validation plus test sets. The red line shows the ideal prediction. } 
\end{figure}

\begin{figure}[hbt]
\centering
\includegraphics[width=0.48\textwidth]{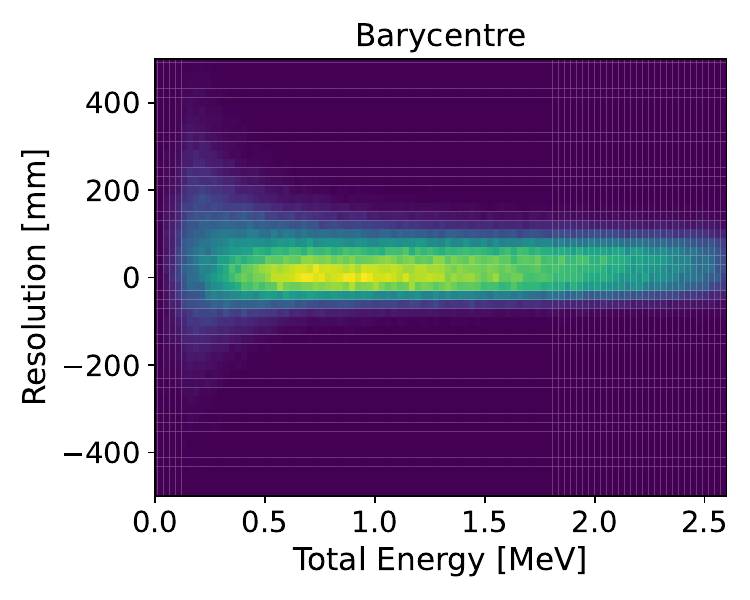}
\includegraphics[width=0.48\textwidth]{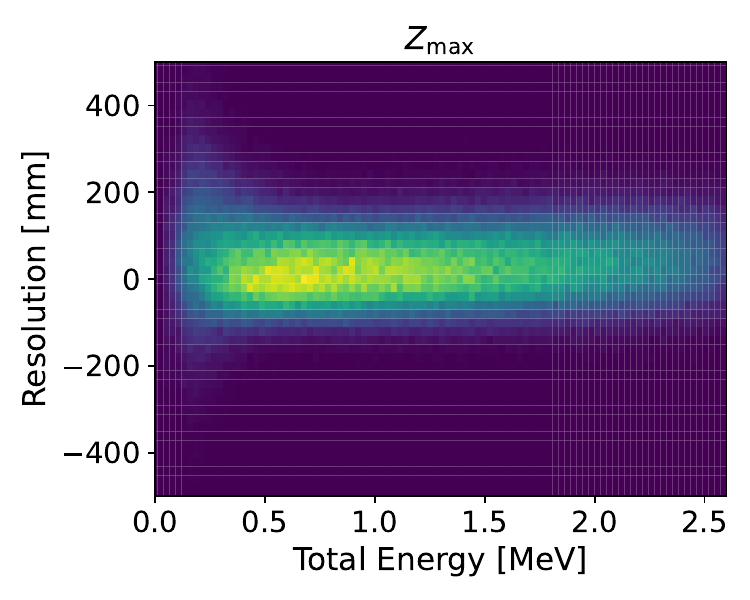}
\caption{\label{fig:CNNZPosRes} CNN resolutions (reconstructed - true) for the (left) barycentre and (right) $z_\textnormal{max}$ as a function of the event energy.  }
\end{figure}

The training included 2.5 M events split into 70/20/10--train/validation/test for single electrons uniformly generated in the energy range 40 keV -- 2.6 MeV. Events are split into individual grouped deposits based on their hit locality, such that x-rays, Bremsstrahlung, and Compton scatters are treated separately. They are then voxelized with 8$\times$8$\times$8~mm$^3$ cube-size and shifted to the centre of the detector based on the event barycentre to remove information about their initial start position, as done in Ref.~\cite{NavarroDiff}. The results of the network on the validation plus test set ($\sim$750 k events) are shown in Fig.~\ref{fig:CNNZPos}, with resolutions as a function of the energy shown in Fig.~\ref{fig:CNNZPosRes}. The test set performance was checked to be in agreement with the validation set. 

The standard deviation of the resolution (reconstructed - true) for the barycentre and $z_\textnormal{max}$ across all event energies is 6.8 cm and 8.5 cm, respectively. This improves to 4.6 cm and 7.5 cm for tracks with an energy greater than 1 MeV. Overall, this precision will be suitable for placing events in 3-D for a fiducial volume cut and calibration in the experiment using the diffusion of the tracks.

\section{Energy: Proportional Avalanche Gain }\label{sec:eres}

Energy resolution at the $\leq$1\% FWHM level is essential for an effective discrimination margin against $2\nu\beta\beta$ and rejection of $\gamma$-ray-induced single-electron backgrounds. At the Q-value for $0\nu\beta\beta$ in $^{136}$Xe, energy is represented in $\simeq$100,000 secondary electrons liberated by the energetic primary electrons and their immediate secondaries.  The energy resolution task is to “count” all $\sim$100,000 secondary electrons with sufficient precision. With the assumption that the Fano factor is $F = 0.15$ in pure xenon gas, the fluctuation in secondary electrons is $\sigma = 125$ electrons. The ``intrinsic'' resolution in the gas phase at the $Q_{\beta\beta}$ energy is: 

\begin{equation}\label{eq1}
  \delta E/E = 0.29\textnormal{\% FWHM},    
\end{equation}
  
\noindent about a factor of four worse than that of germanium diodes~\cite{knoll2010radiation}. 

The use here of proportional avalanche gain may seem highly disadvantageous relative to the almost noise-free gain mechanism of electroluminescence (EL). Yet even with the nearly noise-free gain of EL, significant fluctuations enter from the Gaussian statistics of photon detection and systematic errors enter from substantial variations in optical paths. These contributions dominate the energy resolution in a large EL-based system such as NEXT.  Yet energy resolution provided by proportional avalanche gain can arguably be expected to reach 1\% FWHM, perhaps even reaching 0.8\% FWHM if systematic effects are fully tamed. In the presence of instrumental effects, energy resolution is well described by,

\begin{equation}\label{eqn:eres_fluc}
    (\sigma/N)^2 = (f + F) / N,
\end{equation}

\noindent where $\sigma$ is the root-mean-square (rms) numerical error in counting $N$ electrons. The factor $f$ represents the total impact of the amplification fluctuations and other effects, and $F$ is the Fano factor~\cite{PhysRev.72.26}. 

Ramsey and Agrawal, in two related works aimed at x-ray astronomy, \cite{RamAgra}, ~\cite{ctx44366358300004911} examined selected molecular additives in xenon for optimum energy resolution in proportional avalanche cylindrical and parallel plate chambers. While energy resolution differences were typically less than 10\%, isobutylene or propylene at $\sim$5\% concentration in xenon displayed the best  performance:
\begin{equation}
    \delta E/E = 8\textnormal{\% FWHM},
\end{equation}

\noindent at 22 keV in a cylindrical geometry, even for gains up to 1000. This resolution is consistent with a value of $f=1$ in Eqn.~\ref{eqn:eres_fluc}. Purity variations appeared to be responsible for at least some of the differences among candidates. The tendency for resolution to deteriorate at higher gains in the cylindrical geometry, but not in the parallel plate geometry in which avalanches are more dispersed, is attributed to positive ion space charge effects affecting local electric fields in the avalanche region. Even for gains of 1000, the NTP-TPC is unlikely to experience such adverse effects due to the larger diffusion of the electrons prior to their arrival at the anode plane.

These authors also examined molecular candidates for Penning effects, which would appear as a large reduction in HV for a specific gain relative to other gases. A dramatic effect was observed only for mixtures with TMA or dimethylamine (DMA) --- as much as 55\% less, relative to xenon plus CO$_2$. Reduced HV could be beneficial for electrostatic stability of the anode plane, described  in Section ~\ref{subsec:anode}.  TMA has a measured ionization potential of 7.9 $\pm$ 0.2 eV, apparently overlapping well with the excitation energy of atomic xenon at 8.14 eV, or with xenon excimers.  The presence of a strong Penning effect in xenon + TMA may even lead to a Fano factor smaller than 0.15, a small but beneficial effect for energy resolution~\cite{NygrenEres}.  

The same resolution of 8\% FWHM with the 22 keV $^{109}$Cd x-ray was also obtained using the cylindrical proportional counter filled with 95\% xenon and 5\% TMA~\cite{RamAgra} at a gain of $\sim$300. Extrapolating these 22 keV results  to $Q_{\beta\beta}$, a factor of $\sim$111, a resolution of:

\begin{equation}
    \delta E/E \sim 8/\sqrt{111} = \textnormal{0.76\% FWHM},
\end{equation}

\noindent is predicted for E = $Q_{\beta\beta}$, which is also expected with a value of $f$=1.  Such a huge extrapolation cannot, of course, be taken at face value. Yet, as each electron is multiplied independently of others and with statistically equivalent early fluctuations, such extrapolation is defensible in principle. 

The authors observed a noticeably poorer resolution,$\sim$13\% worse, in xenon plus carbon dioxide mixtures relative to xenon plus propylene or isobutylene, yet it could turn out that if slightly worse energy resolution is tolerable, the virtue of carbon dioxide's inertness may in practice be greater than the benefits of the organic gases.  Xenon plus 5\% propylene does not present a serious flammability hazard, but the additional toxicity concerns for TMA could be a possible complication. However, the high mechanical integrity required of the gas containment for the xenon isotope could provide the needed safety margin for the molecular additives as well. Further R\&D is warranted.

The cadmium x-ray leaves a very short electron track that allows essentially full collection of ionization electrons at the anode well within the shaping time of the electronic signal shaping chain. At 2.5 MeV, however, the ionization electrons are spread out over meters in space and a fraction of a millisecond in drift time, indicating a new level of technical challenge. To account for all poorly known calibration errors and systematic effects, a “constant term” is often added in quadrature. 

The systematic effects that may be present in the proportional avalanche gain design proposed here are numerous and dangerous, possibly degrading practical performance by a large factor.   The energy resolution of the Gotthard experiment ~\cite{PhysRevD.48.1009,Luscher:1998sd} was dominated by some of these effects, leading to an inferred resolution of $\sim$5\% FWHM at 2.5 MeV, a factor of five worse than the optimistic value, $\sim$1\% FWHM suggested above. As our multi-wire anode plane design resembles theirs,  justification for optimism is required. These effects are listed in more detail and partially addressed in Appendix~\ref{appdx:gotthard}; a full accounting is a primary goal of future work. An important part of that work will address the significance of induced signals and required voltage stabilizations for all the anode plane electrodes. 

The secondary ionic signal also carries a level of energy information roughly equivalent to that of the anode plane, perhaps offering some complementary measure of energy with different systematics, or providing support for the $^{83\textnormal{m}}$Kr signals used for calibration. Flooding the detector with $^{83\textnormal{m}}$Kr continuously will be necessary to populate the calibration database with necessary statistical precision.  The $^{83\textnormal{m}}$Kr decay sequence deposits 41.5 keV, yielding $\sim$1700 electrons.  Signal sharing between neighboring anode wires is common but should be a manageable problem.  $\gamma$-rays with known energy near $Q_{\beta\beta}$ will supply the overall scale factor and check of linearity.

As the ratio of the gain region electric field to the drift electric field is large, a very small fraction -- and statistically indifferent to event topology -- of primary electrons will land on the grid wires of the anode plane. This loss has a negligible statistical impact on energy measurement. A virtue of using a wire plane to measure energy is that edge definition and noise summation issues are much smaller relative to a design using pixelation. 

The primary electrons from either $0\nu\beta\beta$ or $\gamma$-ray-induced events gain/lose energy from/to the imposed drift electric field, depending on their trajectory. At 16 kV across 4 m, this energy contribution is just 4 keV/m, an insignificant effect relative to 25 keV, the 1\% FWHM value. There is also a gain effect due to static gas density variations with altitude within the detector, which should be straightforward to calibrate.


\section{Event Topology – Results and Opportunity}\label{sec:topology}

The objective of the NTP-TPC is to utilize the detailed reconstruction of the electron tracks, with a diffusion set near the thermal limit, to reject backgrounds at a sufficient level to reach 0.1 counts/ton/year. The results from a simple analysis with purely algorithmic reconstruction of topological features revealed that a background level reaching 0.4 counts/ton/year is obtainable~\cite{OptimalParamsGXeTPC}, very encouraging even in the absence of fully realistic experimental simulations. 

Here, we explore a new analysis with more advanced machine learning (ML) reconstruction methods, including graphical (GNN) and 3-D sparse convolutional neural (CNN) networks. Each architecture explores alternative approaches to tackling the signal-background classification task. The GNN utilizes physics-derived quantities such as the angle of scatter and identifies track segments corresponding to the primary track, $\delta$-rays, or $\gamma$ interactions. The CNN uses an approach that allows the ML network to learn by providing voxelized energy-weighted hits as 3-D images. Details of the model architectures and algorithms used are given in Appendix~\ref{appdx:ML}. We also train each ML model after applying some loose algorithmic cuts before the ML training, which remove the simplest background events and significantly reduce the overall dataset size while increasing the training speed (labeled ``CNN/GNN+Alg''). The analysis assumes the same experimental assumptions, simulation, and rate calculation\footnote{This analysis increased the event sample in training to $\sim$750k events per event category ($0\nu\beta\beta$, $^{137}$Xe,  $^{214}$Bi, $^{208}$Tl), with a 70/20/10 -- train/validation/test split.} as in Ref.~\cite{OptimalParamsGXeTPC} with assumed energy resolution of 0.75\% FWHM. 

\begin{figure}[hbt]
\centering
\includegraphics[width=0.48\textwidth]{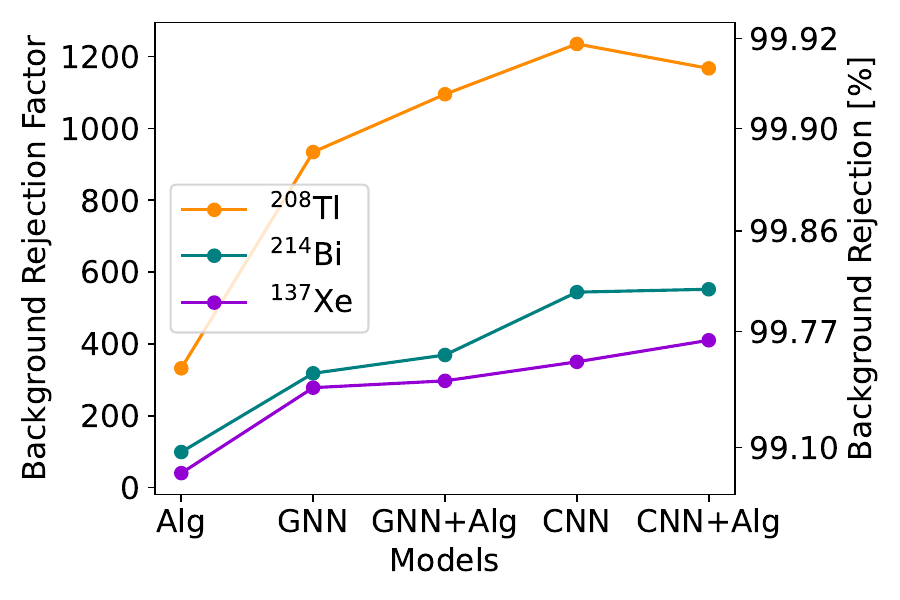}
\includegraphics[width=0.48\textwidth]{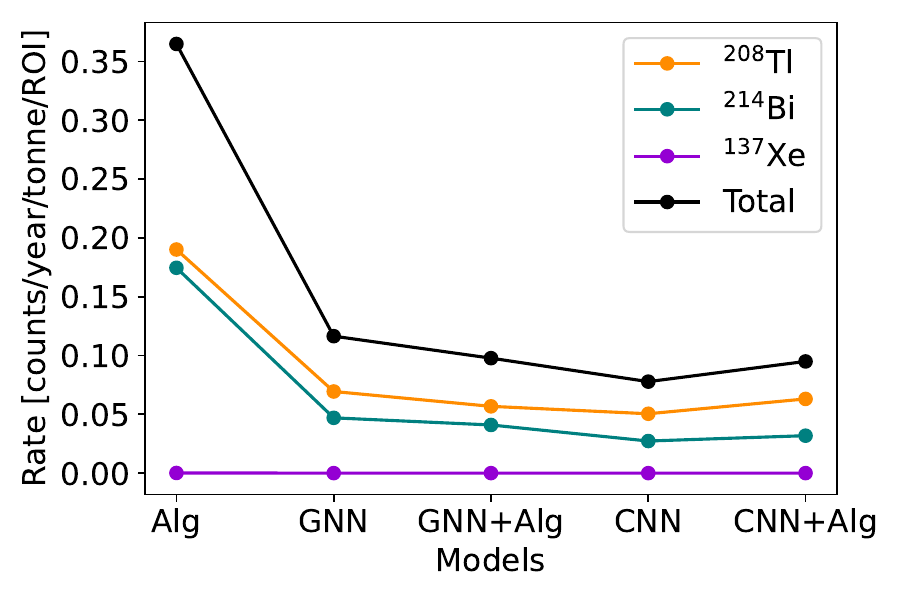}
\caption{\label{fig:MLPerf} (left) Performance of the various neural networks for background rejection compared with a pure algorithmic approach, ``Alg'' from Ref.~\cite{OptimalParamsGXeTPC}. The background rejection factor is defined as the reciprocal of the background efficiency. These performances are for 58\% efficiency at this stage. (right) The total background rate for each model. }
\end{figure}

Figure~\ref{fig:MLPerf} shows the background rejection performance and total background rate for these models compared with a pure algorithmic approach at a signal efficiency of 58\%. After accounting for energy resolution and containment cuts, this results in a final 25\% signal efficiency, which yields about 1 signal count per ton-year at 10$^{27}$ yr half-life. Overall, the ML analyses are almost a factor of four better than a pure algorithmic approach, and each ML model predicted rate is just over or less than the desired rate of 0.1 counts/ton/year/ROI. 

The selected events that present the greatest risk of misclassification characteristically contain a high-energy $\delta$-ray created close to the origin of the primary electron. In this scenario, the ``stub'' associated with the beginning of the primary track is reconstructed as a $\delta$-ray while the $\delta$-ray itself is reconstructed as a second electron creating signal-like topology. 

Further improvements in the reconstruction, perhaps critically important, could be made by utilizing the kinematic angles of $\delta$-ray production and the straightness of the stub. This information could be used to test which stub is the true $\delta$-ray. In addition, for $\gamma$-ray-induced single-electron backgrounds interacting via the photoelectric effect, tagging the $\sim$30 keV xenon K-shell x-ray near the primary electron track start may provide a boost in background rejection performance, with likely a small loss of sensitivity to signal. 

While these considerations indicate the existence of a practical upper limit of discrimination, as there will be events in which the primary stub is too short to reconstruct, we have increasing confidence that the goal of a result at the one-to-several ton-year exposure scale is attainable. 


\section{Demonstrator}\label{sec:demonstrator}

The demonstrator TPC should hold about 100 kg of natural xenon (+ molecular admixture at a few percent level) with an active volume of $\sim$18 m$^3$. The geometry could be either symmetric or asymmetric (single-ended). A symmetric TPC might be better since the geometry more closely mimics the final intended design. On the other hand, the surface-to-volume ratio of an asymmetric TPC would be better.  In any case, the design and performance of the demonstrator must be sufficient to justify a follow-on proposal directly to the ton-scale. Insufficient attention to detail could lead to an unfortunate and false outcome -- that the NTP-TPC approach is not viable. 

The symmetric demonstrator TPC has dimensions 2 $\times$ (2.1 m)$^3$, while an asymmetric cube-style demonstrator TPC has dimensions (2.6 m)$^3$. The HV would not exceed 8 kV. Approximately 264 anode plane channels and 18 panes $\times$ 60 electrodes/pane $\times$ 2 cathode planes = 2160 channels will be required. Some copper shielding may be needed to suppress ambient backgrounds and allow external $\gamma$-sources to dominate the trigger rate. High-quality electronics close to final designs with sufficient sampling depth will be necessary.  Siting the demonstrator at an underground laboratory such as Canfranc, SURF, SNOLAB, or Boulby would be suitable for the demonstrator~\cite{10.3389/fphy.2024.1516502}. 

Using single-electron tracks induced by $\gamma$-rays from the 662 keV $\gamma$-ray of $^{137}$Cs as well as higher energy sources, six essential goals stand out:

\begin{enumerate}
    \item Demonstration of $\leq$1.5\% FWHM energy resolution at $\geq$MeV energies, requiring only a short extrapolation to $Q_{\beta\beta}$. 
    \item “Head-tail” classification performance at $>$99.5\% , with convincing matching  simulations.
    \item 3-D reconstruction using anode plane and cathode plane signals with $>$99.5\% accuracy. 
    \item Demonstration that space charge effects and ionic clustering do not affect the placement in $y$. 
    \item Operational stability using xenon + admixture in an underground setting.
    \item Sufficient overall performance to justify a transition directly to the ton-scale.
\end{enumerate}

Highly collimated MeV  $\gamma$-rays would be introduced at a variety of angles and locations. No credible demonstrator cost estimate can be given now, but it arguably seems modest in the framework of the quest for definitive observations of $0\nu\beta\beta$. Other focused R\&D will be needed, such as optimization of molecular admixtures for energy resolution and a demonstration of operational stability of the several meters-long anode plane wires at the intended voltages.


\section{Perspective and Summary}\label{sec:summary}

Altogether, a plausible scenario based on ``topological supremacy'' seems to be coming into soft focus, perhaps offering a path toward a definitive ton-scale xenon $0\nu\beta\beta$ search. 
Our motivation is to develop a plausibly higher performance path toward discovery in the relatively near term.   The scenario presented here, while audacious, is for the most part conservative technologically.  The list of R\&D activities is not short but generally not technologically limited.  Construction and engineering costs for a room-temperature atmospheric-pressure ton-scale detector are likely smaller than for cryogenic or high-pressure systems. Those will likely have to meet more stringent safety as well as radiopurity requirements since their abilities to exploit intrinsic event characteristics may be less powerful. 

\noindent To summarize:
\begin{itemize}
    \item From an unlikely starting point, these initial explorations show that a ton-scale xenon gas TPC at NTP could be a feasible and competitive technical approach.
    \item An algorithmic approach gets close, with ML providing the margin needed to reach the background target of 0.1 counts/ton/yr/ROI.
    \item Greater transparency to $\gamma$-rays largely compensates for the larger surface area. 
    \item Ordinary copper may suffice, which would enable the desired radiopurity goals.
    \item HV requirements are a factor of 6 smaller than for the same design at 15 bars.
    \item Nearly all electronics are outside the xenon volume, with low power dissipation.
    \item Containment of gas underground at NTP is much safer, easier, and cheaper.
    \item Electron attachment will be much smaller at NTP than at higher pressures.
    \item Canfranc, SURF, SNOLAB, Boulby, or possibly elsewhere, provide natural settings for both demonstrator and ton-scale experiments.
    \item R\&D for a demonstrator is extensive but seems technically straightforward.
    \item A ton-scale search could be undertaken at a pace consistent with other searches. 
\end{itemize}


\acknowledgments

We thank Adam Para (FNAL), Ben Jones (Manchester), and UTA colleagues Jonathan Asaadi and Amir Shamoradi for the many valuable discussions.


\bibliographystyle{JHEP}
\bibliography{biblio}


\appendix
\section{Detailed CSA requirements for electron and ion sensing}
\label{appdx:ion-sensing}

This appendix develops the feasibility criteria behind the cathode-readout choices in Section~\ref{subsec:cathode} and separates them from the much faster anode-electron readout. The detector baseline is fixed by Fig.~\ref{fig:cathode}: a cathode strip plane connected through sealed vias to exterior CSAs. It directly reads secondary ions; chopping or physical modulation coupled to the same strip geometry is the preferred primary-ion development path. A \TM CMOS pixel plane remains a future alternative. These cases occupy different charge, input-capacitance, shaping-time, feedback-memory, and interference regimes; an ENC number without those conditions is not transferable.

\begin{center}
    \small
    \begin{tabular}{p{0.42\textwidth}p{0.42\textwidth}}
        \hline
        Quantity & Baseline value \\
        \hline
        Maximum electron drift time & 3.9~ms ($L=3.93$~m, $v_{\mathrm{e}}\simeq1$~mm/$\upmu$s) \\
        Anode shaping time & Approximately 1--10~$\upmu$s \\
        Ion arrival delay at the cathode & Approximately 10--17~s \\
        Ion signal width and shaping time & Approximately 18--30~ms; test through 100~ms \\
        Ion-CSA feedback decay & Approximately 0.18--1~s for $<5\%$ raw deficit; 0.9--5~s for $<1\%$ \\
        \hline
    \end{tabular}
\end{center}

\subsection{Architectures and charge requirements}

The default NTP-TPC cathode places approximately 8~mm-pitch strip collectors across a 50~cm panel and contacts them through vias to electronics outside the panel. The complete conductor is part of the CSA input. In contrast, \TM uses an exposed patch in the topmost CMOS metal layer as the collection electrode and connects it immediately to circuitry underneath. The \TMS concept extends this arrangement to a tiled TPC plane with electrostatic focusing, local amplification, and on-sensor digitization~\cite{Mei2020TopmetalPlane}; its inter-chip wiring carries power, control, buffered outputs, or digital data rather than extending the sensitive collection node. The relevant architectural comparison is given in Table~\ref{tab:ion-architectures}.

\begin{table}[hbt]
    \centering
    \caption{Direct cathode-readout options. The input capacitance, rather than the nominal segmentation pitch alone, controls the series-noise penalty.}
    \label{tab:ion-architectures}
    \begin{tabular}{p{0.22\textwidth}p{0.31\textwidth}p{0.34\textwidth}}
        \hline
        Option & Sensitive input node & Principal system trade-off \\
        \hline
        Baseline strip CSA & Full panel conductor, via, routing, protection, and CSA gate & Direct secondary-ion readout with exterior electronics; large capacitance, pickup area, leakage paths, and microphonic sensitivity \\
        Primary-ion modulation & The same collectors plus input commutation, balanced field electrodes, or a coupled charge modulator & Preferred primary-ion R\&D path; moves the measurement above the flicker corner but introduces feedthrough and switching or actuator complexity \\
        Future \TM plane & Small exposed on-chip electrode and local CSA; electrostatic focusing covers the larger pitch & Low analog-node capacitance, but replaces the strip plane and adds many dies, in-volume power, networking, cooling, and radiopurity constraints \\
        \hline
    \end{tabular}
\end{table}

For one representative 8~mm topology sample, the assumed primary and secondary charges are
\begin{align}
    Q_{\mathrm{p}} &= N_{\mathrm{p}}q
    \simeq 300q \simeq 0.048~\mathrm{fC}, \\
    Q_{\mathrm{si}} &= Gf_{\mathrm{IBF}}N_{\mathrm{p}}q
    \simeq 7.5\times10^{4}q \simeq 12~\mathrm{fC},
    \label{eq:ion-sample-charge}
\end{align}
where $G=500$ and $f_{\mathrm{IBF}}=0.5$ are the assumed avalanche gain and ion-backflow fraction. If a fraction $f_{k}$ of a sample is assigned to channel $k$, the noise condition for a specified signal-to-noise target $R$ is
\begin{equation}
    \mathrm{ENC}_{k} \leq \frac{f_{k}N}{R}.
    \label{eq:ion-channel-enc}
\end{equation}
Taking $R=5$ and $f_k=1$ only as an illustrative benchmark gives 60~electrons for primary ions and $1.5\times10^{4}$~electrons for secondary ions. This choice is not a detector acceptance criterion. The required $R$ must be obtained from the desired reconstructed-hit efficiency and event-level false-hit probability after matched filtering across all channel--time trials, including non-Gaussian baseline tails. Diffusion and charge sharing make $f_k<1$ and tighten the per-channel limit. A matched filter can combine samples whose waveform is known, but it cannot recover spatial detail that never rises above channel noise and threshold.

Table~\ref{tab:csa-regimes} summarizes the four signal regimes. Here $\tau_{\mathrm{s}}$ is the effective shaping or matched-filter duration and $\tau_{\mathrm{f}}=R_{\mathrm{f}}C_{\mathrm{f}}$ is the CSA feedback-decay constant. They are independent design parameters. A long digital filter cannot restore charge already discharged through a feedback path that is too fast.
\begin{table}[hbt]
    \centering
    \small
    \setlength{\tabcolsep}{3pt}
    \caption{Signal and CSA regimes for one representative topology sample. Published sub-millisecond monolithic filters optimize gainless electron energy measurement; they are not ion-regime noise measurements.}
    \label{tab:csa-regimes}
    \begin{tabular}{p{0.20\textwidth}p{0.14\textwidth}p{0.27\textwidth}p{0.29\textwidth}}
        \hline
        Signal & Charge per sample & Arrival and effective shaping & Principal CSA consequence \\
        \hline
        Anode electrons, $G=500$ & $1.5\times10^{5}q$ (24~fC) & Few-$\upmu$s arrival; $\tau_{\mathrm{s}}\sim1$--10~$\upmu$s to preserve 1~MHz $z$ sampling & Large signal; short-shaping ENC of several thousand electrons is acceptable for energy and topology \\
        Electrons, $G=1$ & $300q$ (0.048~fC) & Few-$\upmu$s arrival; approximately 0.18--1~ms software energy filters demonstrated or simulated for low-rate monolithic planes & Timing and energy may use different filters; electronics noise is important in the summed energy \\
        Cathode secondary ions & $7.5\times10^{4}q$ (12~fC) & Thermal width 18--30~ms; design and test $\tau_{\mathrm{s}}$ up to 100~ms & Large signal, but long $\tau_{\mathrm{f}}$, leakage, flicker noise, and charge sharing must be measured \\
        Cathode primary ions & $300q$ (0.048~fC) & Same 18--30~ms width and up-to-100-ms design range & ENC scale $300f_k/R$ electrons; illustrative $R=5$, $f_k=1$ benchmark is 60~electrons \\
        \hline
    \end{tabular}
\end{table}

\subsection{Input capacitance of an external collector}

For the strip implementation, the first transistor sees:
\begin{equation}
    C_{\mathrm{T}}=C_{\mathrm{col}}+C_{\mathrm{mut}}+C_{\mathrm{via}}+C_{\mathrm{route}}
    +C_{\mathrm{ESD}}+C_{\mathrm{f,in}}+C_{\mathrm{g}}.
    \label{eq:ion-input-capacitance}
\end{equation}
Here $C_{\mathrm{col}}$ denotes capacitance to grounded or AC-grounded structures and $C_{\mathrm{mut}}$ represents the contribution of neighboring electrodes under their actual bias and readout impedances. A single scalar electrostatic estimate is insufficient when neighboring channels move in common mode; the full capacitance matrix and front-end boundary conditions are required.

An 8~mm by 50~cm strip has area $A=40$~cm$^{2}$. If a conducting surface lies behind it at separation $d$, the parallel-plate term alone is:
\begin{equation}
    C_{\mathrm{col}}\simeq\frac{\epsilon_{0}\epsilon_{r}A}{d}
    \simeq35~\mathrm{pF}\,\epsilon_{r}\left(\frac{1~\mathrm{mm}}{d}\right).
    \label{eq:ion-strip-capacitance}
\end{equation}
A 1~mm polymer layer with $\epsilon_r=2$--3 would consequently contribute approximately 70--110~pF. A pane without a nearby backing conductor may be substantially lower, but support frames, neighboring strips, vias, protection, and the MOS gate remain. Until a field model and full-size measurement exist, 10, 30, 50, and 100~pF are useful test loads.

The primary charge expressed as an input-node voltage illustrates the scale of the comparison is shown in Table~\ref{tab:ion-voltage-requirement}.
\begin{table}[hbt]
    \centering
    \caption{Equivalent voltage for 300 primary ions and the total rms input-voltage-noise ceiling for the illustrative $R=5$, $f_k=1$ benchmark.}
    \label{tab:ion-voltage-requirement}
    \begin{tabular}{c c c}
        \hline
        $C_{\mathrm{T}}$ & $Q_{\mathrm{p}}/C_{\mathrm{T}}$ & Allowed rms noise \\
        \hline
        10~pF  & 4.81~$\upmu$V & 0.96~$\upmu$V \\
        30~pF  & 1.60~$\upmu$V & 0.32~$\upmu$V \\
        50~pF  & 0.96~$\upmu$V & 0.19~$\upmu$V \\
        100~pF & 0.48~$\upmu$V & 0.096~$\upmu$V \\
        \hline
    \end{tabular}
\end{table}
These are integrated voltage-noise limits over the signal estimator, not spectral densities at a single frequency.

\subsection{CMOS CSA noise and shaping-time optimization}

Let the input-device voltage-noise spectrum be $S_e(f)=e_{\mathrm{w}}^{2}+A_f/f$, and let $i_{\mathrm{w}}$ denote the aggregate white parallel-current noise from the collector, gate, protection structures, bias network, and feedback or reset element. For a geometrically scaled family of shapers, a useful model is~\cite{Sansen1990NoiseLimits,OConnor2002ScaledCMOS}
\begin{equation}
    \mathrm{ENC}^{2}(\tau) \simeq
    a_{\mathrm{s}}\left(\frac{C_{\mathrm{T}}e_{\mathrm{w}}}{q}\right)^{2}\frac{1}{\tau}
    +a_{\mathrm{f}}\left(\frac{C_{\mathrm{T}}}{q}\right)^{2}A_f
    +a_{\mathrm{p}}\left(\frac{i_{\mathrm{w}}}{q}\right)^{2}\tau
    +\mathrm{ENC}_{\mathrm{ADC}}^{2}+\mathrm{ENC}_{\mathrm{pickup}}^{2}.
    \label{eq:ion-enc}
\end{equation}
The dimensionless coefficients $a_i$ depend on the filter definition. For a fixed input transistor, the white-series ENC scales as $C_{\mathrm{T}}/\sqrt{\tau}$, whereas the flicker ENC scales approximately linearly with $C_{\mathrm{T}}$ and is nearly independent of the absolute shaping time. Parallel-current ENC grows as $\sqrt{\tau}$. Increasing the integration time therefore suppresses only one of the three principal terms.

Ignoring flicker and fixed readout terms, minimizing the sum $A/\tau+B\tau$ gives
\begin{equation}
    \tau_{\mathrm{opt}}=\sqrt{\frac{a_{\mathrm{s}}}{a_{\mathrm{p}}}}
    \frac{C_{\mathrm{T}}e_{\mathrm{w}}}{i_{\mathrm{w}}},
    \qquad
    \mathrm{ENC}_{\mathrm{white,min}}^{2}=2\sqrt{AB}.
    \label{eq:ion-optimal-shaping}
\end{equation}
Thus, a larger detector capacitance moves the optimum toward longer shaping unless current noise is also reduced, but the flicker term remains. Random-telegraph fluctuations, dielectric polarization, temperature drift, microphonics, and coherent pickup are not captured adequately by this stationary three-term expression and can dominate a millisecond-to-second measurement~\cite{Jakobson1997CSA,Ratti2009CMOSCSA}.

The input transistor can be resized for the load. Its approximate thermal voltage-noise density scales as $e_{\mathrm{w}}^{2}\propto1/g_m$, while a simple long-channel flicker model scales as $A_f\propto1/(C_{\mathrm{ox}}^{2}WL)$. Increasing $WL$ therefore reduces gate-referred flicker noise but increases $C_{\mathrm{g}}\simeq C_{\mathrm{ox}}WL$, which itself enters $C_{\mathrm{T}}$. In a flicker-dominated weak-inversion optimum, the gate capacitance is of the same order as detector-plus-parasitic capacitance. Under ideal resizing the minimum flicker ENC can grow more slowly than linearly, approximately as the square root of detector capacitance, but the required device becomes large and costs die area, bias current, stability margin, and additional layout parasitics. There is consequently no process-independent ``intrinsic CMOS ENC.'' A foundry-specific transistor model and extracted layout are necessary.

\paragraph{Concrete SKY130 example.}
As a reproducible process-specific illustration, we evaluated the open-source Skywater 130nm CMOS process, TT 1.8~V PMOS model~\cite{SkyWaterPDK}. The model gives $t_{\mathrm{ox,e}}=4.23$~nm, hence $C_{\mathrm{ox}}=8.16$~fF/$\upmu$m$^2$, and uses the BSIM number-fluctuation model with $N_{\mathrm{OIA}}=1.5\times10^{42}$, $N_{\mathrm{OIB}}=N_{\mathrm{OIC}}=0$, $E_F=1$, and $\texttt{fnoimod}=1$. A transparent long-channel reduction of that model~\cite{Liu2000BSIM4} is
\begin{equation}
    S_{v,1/f}(f)\simeq
    \frac{q^2 kT N_{\mathrm{OIA}}}
    {\alpha C_{\mathrm{ox}}^2WL f},
    \qquad
    S_{v,\mathrm{w}}\simeq\frac{4kT\gamma}{g_m},
    \label{eq:sky130-input-noise}
\end{equation}
where $\alpha=10^{10}$~m$^{-1}$ is the normalization factor appearing in the BSIM implementation and we use the model's $r_{\mathrm{noia}}=0.69$ as the equation-level thermal coefficient $\gamma$. For equal-time-constant CR--RC shaping sampled at its peak,
\begin{equation}
    \mathrm{ENC}_{\mathrm{w}}
    =\frac{\mathrm{e}C_{\mathrm{T}}}{q}
    \sqrt{\frac{S_{v,\mathrm{w}}}{8\tau}},
    \qquad
    \mathrm{ENC}_{1/f}
    =\frac{\mathrm{e}C_{\mathrm{T}}}{q}
    \sqrt{\frac{A_f}{2}}.
    \label{eq:sky130-crrc-enc}
\end{equation}
Taking $C_g\simeq(2/3)C_{\mathrm{ox}}WL$, minimization of the flicker term gives $C_g=C_d$. Table~\ref{tab:sky130-noise-estimate} uses that optimum, $I_D=100~\upmu$A, and $g_m/I_D=20$~V$^{-1}$. It excludes every capacitance other than $C_d+C_g$ and is therefore optimistic.

\begin{table}[hbt]
    \centering
    \small
    \caption{SKY130 TT input-transistor-only ENC estimate. Values separated by slashes use 18, 30, and 100~ms CR--RC shaping. The compact-model column integrates the ngspice input-referred spectrum for parallel 90~$\upmu$m by 1~$\upmu$m PMOS devices nearest the equation-optimum area.}
    \label{tab:sky130-noise-estimate}
    \begin{tabular}{c r r r r r}
        \hline
        $C_d$ (pF) & $WL$ ($\upmu$m$^2$) & $1/f$ ENC & equation ENC & PDK $C_{gg}$ (pF) & BSIM ENC \\
        \hline
        10 & 1837 & 87 & 87/87/87 & 6.9 & 71/71/71 \\
        30 & 5512 & 150 & 150/150/150 & 18.2 & 129/129/129 \\
        50 & 9187 & 194 & 194/194/194 & 28.4 & 164/164/164 \\
        100 & 18375 & 274 & 275/274/274 & 52.1 & 222/221/221 \\
        \hline
    \end{tabular}
\end{table}

Equation~\ref{eq:ion-channel-enc} gives the comparison scale $300f_k/R$~electrons for a primary-ion sample. For the explicitly illustrative $R=5$, $f_k=1$ case, both transistor-only estimates exceed 60~electrons throughout the 10--100~pF range. For $R=3$, the ceiling is 100~electrons and the optimistic 10~pF cases fall below it, while the 30--100~pF cases remain above it. The weak shaping-time dependence is the expected signature of flicker dominance. The equation and BSIM values need to agree only at order-unity level because the former omits bias dependence, effective noise dimensions (including \texttt{lintnoi}), and channel-length-modulation and correlation terms. Neither calculation includes extracted-layout parasitics, feedback and leakage-current noise, reset artifacts, RTS, pickup, or process spread. SKY130 therefore quantifies margin versus capacitance and $R$; it does not set detector acceptance, predict a completed CSA, or establish a general CMOS limit.

For an optimistic leakage-only estimate, shot noise gives
\begin{equation}
    \mathrm{ENC}_{\mathrm{par}}\simeq\sqrt{\frac{2I_{\mathrm{leak}}\tau}{q}}.
    \label{eq:ion-parallel-noise}
\end{equation}
Assigning the entire illustrative 60-electron benchmark to this term gives the upper limits in Table~\ref{tab:ion-leakage}. For a selected $R$ and $f_k$, multiply these tabulated limits by $(5f_k/R)^2$; the actual leakage allowance must be lower because series, flicker, digitization, and pickup noise consume part of the same noise allocation.
\begin{table}[hbt]
    \centering
    \caption{Aggregate leakage-current ceiling if parallel shot noise alone were allowed to reach the illustrative 60-electron ENC benchmark.}
    \label{tab:ion-leakage}
    \begin{tabular}{c c}
        \hline
        Effective shaping time & $I_{\mathrm{leak,max}}$ \\
        \hline
        10~ms  & 28.8~fA \\
        20~ms  & 14.4~fA \\
        50~ms  & 5.8~fA \\
        100~ms & 2.9~fA \\
        \hline
    \end{tabular}
\end{table}
Reset noise and charge injection must also be treated explicitly. A periodically reset feedback capacitor can introduce a $kT/C$ baseline uncertainty and a signal-dependent dead interval. Correlated baseline subtraction can reject a repeatable reset offset, but not leakage fluctuations, random-telegraph steps, or charge injected after the baseline sample. Continuous resistive or MOS feedback avoids a hard reset but adds parallel noise and may have a poorly controlled time constant at the required resistance.

\subsection{Ion arrival, shaping time, and CSA feedback memory}

For ion mobility $\mu$, drift field $E$, and distance $L$, the nominal transport delay is
\begin{equation}
    t_{\mathrm{d}}=\frac{L}{\mu E}.
    \label{eq:ion-drift-delay}
\end{equation}
With $L=4$~m, $E=40$~V/cm, and $\mu=0.6$--1.0~cm$^{2}$/V/s, $t_{\mathrm{d}}$ is approximately 10--17~s. This delay determines how long the electronics and baseline must remain observable, but it is not the shaping time of one arrival feature.

In the thermal limit, the Einstein relation gives $D_L\simeq\mu k_{\mathrm{B}}T/q$. The corresponding longitudinal and temporal widths are
\begin{equation}
    \sigma_z\simeq\sqrt{2D_L\,t_{\mathrm{d}}},
    \qquad
    \sigma_t\simeq\frac{\sigma_z}{\mu E}.
    \label{eq:ion-time-spread}
\end{equation}
For the illustrative parameters above, the thermal estimate gives a longitudinal width near 0.7~cm and a temporal width of approximately 18--30~ms. The ion CSA must consequently support an effective shaping or matched-filter time $\tau_{\mathrm{s}}\simeq18$--30~ms, with a conservative design and test range extending to 100~ms for track extent, multiple species, clustering, charge exchange, field nonuniformity, and the cathode weighting potential. By comparison, the avalanche-electron anode receives local charge in a few $\upmu$s and needs order-1--10-$\upmu$s shaping to preserve the proposed 1~MHz sampling. The ion front end is therefore a distinct low-frequency design.

Its analog memory must also be long. For a CSA with feedback decay $\tau_{\mathrm{f}}=R_{\mathrm{f}}C_{\mathrm{f}}$ receiving constant current for duration $T$, the output at the end of collection relative to the ideal retained-charge step is
\begin{equation}
    \eta(T)=\frac{V(T)}{Q/C_{\mathrm{f}}}
    =\frac{\tau_{\mathrm{f}}}{T}
    \left(1-e^{-T/\tau_{\mathrm{f}}}\right).
    \label{eq:ion-ballistic-deficit}
\end{equation}
Thus $\tau_{\mathrm{f}}=10T$ retains approximately 95.2\% of the raw amplitude, whereas $\tau_{\mathrm{f}}=50T$ retains approximately 99.0\%. Across $T=18$--100~ms, less than 5\% uncorrected ballistic deficit calls for $\tau_{\mathrm{f}}\gtrsim0.18$--1~s, and less than 1\% calls for approximately 0.9--5~s. Digital pole-zero correction can recover a known average decay, but then the feedback constant, baseline, and current noise must remain stable enough that the correction itself does not create a resolution term. A merely ``many-millisecond'' CSA is not automatically adequate for a 100~ms ion waveform.

The instantaneous current on cathode channel $k$ follows the Shockley--Ramo relation~\cite{Ramo1939}
\begin{equation}
    i_k(t)=q\,\mathbf{v}(t)\cdot\mathbf{E}_{\mathrm{w},k}\!\left[\mathbf{x}(t)\right],
    \label{eq:ion-ramo-current}
\end{equation}
where $\mathbf{E}_{\mathrm{w},k}$ is that channel's weighting field. The rectangular pulse used in Eq.~\ref{eq:ion-ballistic-deficit} is only a transparent retention test; the design input must be a field-derived current template. For colored noise, the achievable matched-filter signal-to-noise is determined by the spectral overlap
\begin{equation}
    \mathrm{SNR}^{2}\propto\int_{0}^{\infty}
    \frac{|I_{\mathrm{s}}(f)|^{2}}{S_i(f)}\,df,
    \label{eq:ion-matched-filter}
\end{equation}
with normalization set by the one- or two-sided spectral convention. This calculation separates the several-second transport delay from the 18--100~ms bandwidth relevant to an individual ion-arrival feature.

\subsection{Interpretation of existing CMOS results}

Table~\ref{tab:csa-benchmarks} reports each ENC together with the input capacitance and filter time used to obtain it. The feedback decay is also stated where the cited result provides it.
\begin{table}[hbt]
    \centering
    \small
    \setlength{\tabcolsep}{3pt}
    \caption{CMOS CSA benchmarks. ``Measured'' and ``simulation goal'' are kept distinct; none of these is an ENC measurement with a 50~cm collector and 18--100~ms ion shaping.}
    \label{tab:csa-benchmarks}
    \begin{tabular}{p{0.19\textwidth}p{0.16\textwidth}p{0.26\textwidth}p{0.28\textwidth}}
        \hline
        Front end and status & Input capacitance & Filter and feedback timing & ENC and scope \\
        \hline
        \TMIIm, measured~\cite{An2016TopmetalII} & $C_{\mathrm{T}}\simeq23$~fF & 1~ms FWHM digital trapezoid; $\tau_{\mathrm{f}}=7.6$~ms for the reported test & 13.9~electrons; exposed on-chip collector, with injection capacitance obtained from simulation \\
        \TMS concept, simulation~\cite{Mei2020TopmetalPlane} & $C_{\mathrm{d}}\simeq5$~pF for a 1~mm electrode & Noise minimum at approximately 180~$\upmu$s digital trapezoidal shaping; illustrated response uses $\tau_{\mathrm{f}}=1$~ms & Design goal below 30~electrons, not a measured ENC; ion operation was proposed by retuning $\tau_{\mathrm{f}}$ into many milliseconds \\
        ALICE TPC PASA, measured~\cite{Soltveit2012ALICEPASA} & 25~pF & 160~ns peaking time (190~ns FWHM) & $244+17C[\mathrm{pF}]$, approximately 669~electrons at 25~pF \\
        Yun et al., measured~\cite{Yun2011LargeCapCSA} & 50~pF detector, 10~pA leakage & 90~$\upmu$s shaper time constant & 950~electrons; measured capacitance slope 18~electrons/pF \\
        LTARS TPC ASIC, measured~\cite{Kishishita2020LTARS} & 300~pF & 1~$\upmu$s shaping & $2695\pm71$~electrons \\
        \hline
    \end{tabular}
\end{table}

These are positive CMOS results in their respective regimes, not fundamental limits. \TMIIm establishes very low noise for a monolithic 23~fF node at 1~ms shaping; the \TMS concept gives a credible low-noise design path at 5~pF and 180~$\upmu$s shaping. They support \TM as a future replacement-plane R\&D option, not as the cathode assumed in Fig.~\ref{fig:cathode}. Neither value is transferable to the default 50~cm strip collector or to 18--100~ms ion shaping. Conversely, the higher ENC of a strip-loaded TPC front end says nothing adverse about the intrinsic capability of a monolithic \TM node.

The 50~pF study also calculated an unconstrained 83-electron optimum at 11~$\upmu$s, but only with 22~mA input current and a 33.6~mm-wide transistor; its 50~pF weak-inversion optimum was 348~electrons at 124~$\upmu$s~\cite{Yun2011LargeCapCSA}. Even these calculated values do not address an 18--100~ms ion filter. Achieving the illustrative 60-electron \emph{system} ENC benchmark with tens of picofarads, a 0.2--5~s feedback-decay design range, few-fA effective leakage, and realistic pickup would be a new result. The actual target remains $300f_k/R$~electrons and must be set by reconstruction performance.

\subsection{CSA contribution to the energy resolution}

The avalanche-electron energy measurement is much easier than either gainless electron or primary-ion sensing. At $Q_{\beta\beta}$, let $N_0\simeq10^{5}$ be the total primary ionization electrons and let the reconstructed energy sum contain $M$ independent channel-time estimates with equal input-referred $\mathrm{ENC}_{\mathrm{e}}$. The electronics contribution is approximately
\begin{equation}
    \left(\frac{\Delta E}{E}\right)_{\mathrm{CSA,FWHM}}
    \simeq
    2.355\frac{\sqrt{M}\,\mathrm{ENC}_{\mathrm{e}}}{G N_0}.
    \label{eq:csa-energy-resolution}
\end{equation}
For the representative $M=300$ and $G=500$, each sample contains approximately $1.5\times10^{5}$ avalanche electrons. Even $\mathrm{ENC}_{\mathrm{e}}=2700$~electrons, the measured LTARS value at 300~pF and 1~$\upmu$s shaping, gives per-sample S/N~$\simeq56$ and only 0.22\% FWHM in the summed energy. The 669- and 950-electron benchmarks in Table~\ref{tab:csa-benchmarks} would correspond to 0.055\% and 0.078\% FWHM, respectively. Requiring the CSA term to remain below 0.20\% FWHM permits approximately 2500~electrons ENC per independent estimate. Thus a short-shaping anode CSA need not approach \TM noise: combining a 0.22\% electronics term in quadrature with the manuscript's 0.76\% avalanche-statistics estimate changes it only to approximately 0.79\% FWHM.

This margin relies on the avalanche occurring before the CSA and on digitizing short, few-$\upmu$s pulses for a later event-wide sum; the front end need not hold the complete fraction-of-a-millisecond track as one analog pulse. At $G=1$, Eq.~\ref{eq:csa-energy-resolution} instead gives 0.57\% FWHM for 14-electron ENC and 1.22\% FWHM for 30-electron ENC at $M=300$. Gainless electron sensing is therefore genuinely noise-critical even though its charge arrives quickly. Coherent baseline motion, channel-gain error, saturation, and nonlinearity do not average as $\sqrt{M}$ and must still be controlled separately.

For the cathode secondary-ion sum, the corresponding gain factor is $f_{\mathrm{IBF}}G=250$. If, purely as a design example, a long-shaping ion prototype achieved 3000-electron ENC, its uncorrelated electronics term would be approximately 0.49\% FWHM and its unshared per-sample S/N would be about 25. This value must be measured at 18--100~ms and is not inferred from the short-shaping benchmarks. The primary ions have neither $G$ nor $f_{\mathrm{IBF}}G$ amplification; their per-channel ENC scale is instead $300f_k/R$~electrons, with 60~electrons corresponding only to the illustrative unshared $R=5$ case. The anode avalanche-electron CSA can therefore meet the energy-resolution goal without being the same circuit as the long-memory ion CSA.

\subsection{Chopping and physical signal modulation}

The primary-ion R\&D baseline preserves the Fig.~\ref{fig:cathode} strip collectors and seeks to up-convert their signal before the first noisy transistor. Chopper stabilization after the first CSA cannot recover signal-to-noise already lost in its input device. Electronic chopping is relevant only if the strip-node charge is commutated before or within that stage~\cite{Enz1996Chopper}. The switches add junction and gate capacitance, off-state leakage, charge injection, clock feedthrough, and dead time. Autozeroing and correlated double sampling can remove sampled offset and part of the low-frequency noise, but alias wideband noise into baseband and do not remove all random-telegraph or between-sample drift. These approaches must be tested with the full strip capacitance.

A physical alternative is to modulate the approaching ions electrostatically. Balanced guide electrodes near a collector can add a sinusoidal transverse field $E_{\mathrm{ac}}$ at angular frequency $\omega_{\mathrm{m}}$. In the mobility regime, the ion displacement amplitude is
\begin{equation}
    x_{\mathrm{ac}}\simeq\frac{\mu E_{\mathrm{ac}}}{\omega_{\mathrm{m}}}
    =\frac{\mu E_{\mathrm{ac}}}{2\pi f_{\mathrm{m}}}.
    \label{eq:ion-modulation-amplitude}
\end{equation}
For a locally linear weighting potential $\phi_{\mathrm{w}}$, the corresponding induced-current amplitude is approximately
\begin{equation}
    i_{\mathrm{ac}}\simeq Q\,\omega_{\mathrm{m}}x_{\mathrm{ac}}
    \left|\frac{\partial\phi_{\mathrm{w}}}{\partial x}\right|
    \simeq Q\mu E_{\mathrm{ac}}
    \left|\frac{\partial\phi_{\mathrm{w}}}{\partial x}\right|.
    \label{eq:ion-modulated-current}
\end{equation}
The ideal low-frequency mobility model makes this current amplitude weakly dependent on modulation frequency, while the physical displacement decreases as $1/f_{\mathrm{m}}$. The guide pitch, field gradient, available voltage, and onset of trajectory distortion therefore set an upper useful frequency. Direct capacitive drive pickup, approximately proportional to $2\pi f_{\mathrm{m}}C_{\mathrm{drive}}V_{\mathrm{ac}}$, grows in relative importance. Paired guide electrodes and a differential sense geometry are required so that this much larger coherent feedthrough cancels before synchronous detection.

\begin{figure}[hbt]
    \centering
    \includegraphics[width=0.50\textwidth]{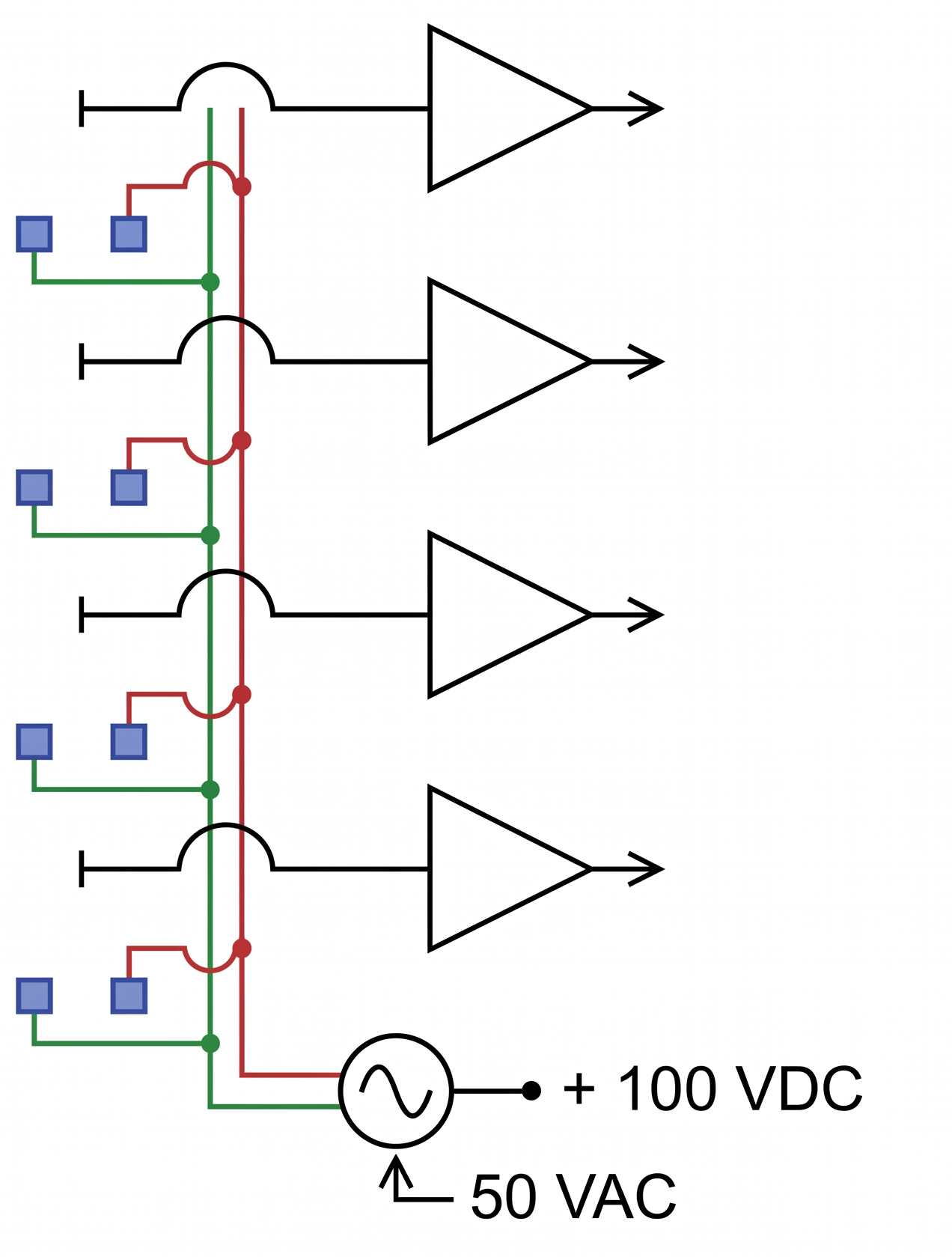}
    \caption{Conceptual ion-field modulation near the cathode. Paired guide electrodes receive the same AC component superimposed on their DC bias. Symmetric coupling to adjacent collectors is intended to reject direct drive pickup while retaining the weighting-field signal from ion motion.}
    \label{fig:ion-modulation}
\end{figure}

\subsection{MEMS capture followed by charge modulation}

The ions need not themselves follow a kilohertz modulation field. They can first be focused onto an electrically isolated conductive island and neutralized, changing the island charge by $Q$. Coupling the island to a driven MEMS variable capacitor translates the static charge into a narrow-band electrical signal. If the island has total capacitance $C_{\Sigma}=C_{\mathrm{p}}+C_0$ and the driven capacitance is $C_0+\Delta C\cos\omega_{\mathrm{m}}t$, then
\begin{equation}
    V(t)=\frac{Q}{C_{\Sigma}+\Delta C\cos\omega_{\mathrm{m}}t}
    \simeq\frac{Q}{C_{\Sigma}}
    -\frac{Q\Delta C}{C_{\Sigma}^{2}}\cos\omega_{\mathrm{m}}t.
    \label{eq:mems-charge-modulation}
\end{equation}
The carrier amplitude is therefore
\begin{equation}
    V_{\mathrm{m}}\simeq\frac{Q}{C_{\Sigma}}
    \frac{\Delta C}{C_{\Sigma}}.
    \label{eq:mems-signal-amplitude}
\end{equation}
This form exposes two design requirements: minimize parasitic capacitance $C_{\mathrm{p}}$ and maximize the safe fractional modulation $\Delta C/C_{\Sigma}$. A narrow-band readout of bandwidth $B$ has approximate voltage S/N $V_{\mathrm{m}}/\sqrt{S_v(f_{\mathrm{m}})B}$, so the carrier should lie above the transistor flicker corner but below frequencies where mechanical displacement, capacitive loss, or drive feedthrough becomes unfavorable.

Micromechanical variable-capacitance electrometers have demonstrated charge-to-AC up-conversion at room temperature and atmospheric pressure~\cite{Riehl2003MEMSElectrometer}; an external charge-transfer electrode has also been coupled to such a device~\cite{Menzel2011Micromechanical}. These devices establish the direct variable-capacitance route.

A complementary MEMS architecture translates charge into a resonance-frequency shift rather than directly reading the voltage in Eq.~\ref{eq:mems-signal-amplitude}. Wang et al. coupled two nonlinear silicon double-ended tuning-fork resonators and operated them near an internal-resonance transition~\cite{Wang2020NonlinearMEMS}. The peak frequency of the driven resonator varied with the electrostatic coupling voltage and was tracked continuously with a phase-locked loop. In their calibration,
\begin{equation}
    \Delta f=K_{\mathrm{sf}}\frac{\Delta Q}{C_{\mathrm{c}}},
    \label{eq:mems-frequency-transduction}
\end{equation}
with $C_{\mathrm{c}}=12.97$~fF and $K_{\mathrm{sf}}=40.60$~Hz/V, corresponding to approximately 0.50~mHz per electron. At room temperature they reported an Allan-deviation charge resolution of $2.1\pm0.9$~electrons, a charge-noise density of $0.197\pm0.056$~electrons/$\sqrt{\mathrm{Hz}}$, real-time closed-loop tracking, and a calibrated dynamic range of approximately $10^{6}$~electrons. Nonlinear coupling therefore offers three useful elements for the present R\&D: enhanced charge-to-frequency transduction, a narrow-band frequency output, and a two-resonator geometry compatible with differential rejection of common-mode feedthrough.

These figures are an existence proof for the transducer, not a demonstrated ion detector. The input charge was emulated by voltage steps on the coupling capacitor rather than deposited and retained on an isolated island; the device operated below 3~Pa in vacuum, used external drive and lock-in/PLL instrumentation, and required active feedthrough cancellation. Operation in xenon requires a new mechanical design. Gas damping lowers the mechanical quality factor $Q_{\mathrm{m}}$, reducing resonant displacement for a given drive but broadening bandwidth and shortening settling time. Brownian force noise, displacement-readout noise, and coherent actuator feedthrough must be included at the chosen mode. A symmetric structure can place the charge signal at the second harmonic while rejecting first-harmonic drive pickup, at the cost of smaller transduction. A TPC prototype must repeat the measurement with known deposited charge, establish a reproducible one-elementary-charge island change per captured positive ion, and operate in atmospheric-pressure xenon under the intended gas-mixture and electric-field conditions. The relevant comparison must report charge-to-frequency gain, Allan deviation, bandwidth, capture efficiency, and false counts under the intended pressure and fields.

Charge must remain on the island throughout capture and measurement. Its leakage time constant $R_{\mathrm{leak}}C_{\Sigma}$ must comfortably exceed the ion-arrival window and synchronous integration time. Surface films, dielectric charging, adsorbed gas, field emission, and the reset switch can all dominate over bulk insulation. Resetting introduces its own $kT/C$ uncertainty and charge injection, so each measurement should include a pre-capture baseline, a post-capture modulated measurement, and a controlled reset verification. The ion-focusing efficiency and probability that neutralization produces exactly one elementary-charge change per incident positive ion must also be established.

\subsection{Prototype sequence}

The R\&D should preserve the strip-plane secondary-ion baseline while developing primary-ion up-conversion on that geometry:
\begin{enumerate}
    \item Calculate the cathode capacitance matrix and weighting fields for the full pane, including supports, vias, neighboring channels, and electronics boundary conditions. Measure a full-size 8~mm by 50~cm collector rather than extrapolating from a short coupon.
    \item Generate primary- and secondary-ion current templates using the actual Xe/additive mobility species, diffusion, charge exchange, track extent, and Shockley--Ramo weighting fields.
    \item Prototype the avalanche-electron anode CSA with the measured full-length wire capacitance. Inject 24~fC, few-$\upmu$s sample pulses at 1--10~$\upmu$s shaping and verify ENC below approximately 2500~electrons, sub-percent channel calibration, linear event-wide summation, and stable pole-zero correction.
    \item Test the default strip-plane ion CSA with 10, 30, 50, and 100~pF loads and programmable $\tau_{\mathrm{f}}$ spanning approximately 0.2--5~s. Inject the 12~fC secondary-ion template at 18, 30, and 100~ms widths and verify the required topology margin. Inject 0.048~fC as a diagnostic of unmodulated primary-ion reach, not as a baseline requirement.
    \item On the same collector load and geometry, compare input-node chopping/commutation, balanced ion-field modulation, MEMS variable-capacitance modulation, and nonlinear-resonator frequency transduction using the 0.048~fC template and measured switching or drive-feedthrough budgets. For either MEMS path, deposit known charge on the capture island rather than substituting only a calibrated voltage step.
    \item For every prototype, report ENC together with total input capacitance, filter definition and duration, feedback-decay constant, leakage, and power. Measure spectra from millihertz to megahertz, Allan deviation, random-telegraph-step distributions, reset artifacts, common-mode rejection, and matched-filter ENC versus duration; for a resonant MEMS readout also report charge-to-frequency gain, mechanical quality factor, tracking bandwidth, and false-transition rate. Repeat across temperature, humidity before sealing, bias history, and representative electromagnetic and mechanical disturbances.
    \item Retain the strip-plane CSA as the detector baseline once it resolves the shared secondary-ion waveform. For primary-ion imaging, report reconstructed-hit efficiency and event-level false hits versus threshold for the shared 0.048~fC template under realistic pickup and stability conditions. Include $R=5$ as a comparison point, not as the acceptance criterion.
    \item As a separate future work package, test a \TM-like tile at its native on-chip capacitance and with appropriate electrostatic focusing. It would replace the cathode strip plane and must not be used as the noise benchmark for that plane.
\end{enumerate}

This sequence keeps the Fig.~\ref{fig:cathode} strip--CSA system as the installed ion readout, makes chopping or physical modulation its primary-ion extension, and preserves \TM as a distinct future alternative without conflating the two input capacitances.


\section{Energy resolution in multi-anode planar geometries}\label{appdx:gotthard}
To aspire to energy resolutions approaching $\sim$0.7 – 1.0\% FWHM for electrons at 2.5 MeV with proportional gain in a multi-anode geometry, all factors that degrade energy resolution must be clearly understood quantitatively and kept under control to a high level. Examples in the literature include Ref.~\cite{BOLOTNIKOV1997360}, which reports a resolution of 0.52\% FWHM using 662 keV gammas with an ionization chamber, supporting a Fano factor of 0.15. The Gotthard TPC experiment, which is an example in a multi-anode geometry~\cite{PhysRevD.48.1009}, reports 6.6\% FWHM at 1.592 MeV. In the absence of systematic effects, a resolution of 5.3\% FWHM at 2.5 MeV may be inferred. This value, a sobering result in a real experiment, suggests that the path toward a demonstration of 1\% FWHM at 2.5 MeV will require very careful, thorough effort.

One of the Gotthard TPC authors (Wong, Ref.~\cite{wong1991search}; p28) refers to Ref.~\cite{allison1976ionisation2} as a guide for energy resolution possible in practice, in an MWPC geometry similar to theirs: “From measurements using a Multiple Wire Proportional Chamber (MWPC) \cite{allison1976ionisation2}, with comparable wire spacings and lengths to the TPC, the main contribution to the energy resolution is caused by the gain variation between different wires ($\sim$4\%). Therefore, we choose 5\% at 2.5 MeV as the goal for the energy resolution of the TPC.” Their relatively modest resolution goal was thus well met in practice. 

The signal derived from an anode wire is the sum of arriving ``fast'' electrons produced in the avalanche plus departing ``slow'' ions. The contribution of the electron component is very small relative to that of the ions. In the ``ideal'' geometry of the cylindrical proportional counter, the integrated signal on the anode is, by symmetry,  indifferent to azimuthal variations in the incoming primary electron ensemble.   However, in a multi-wire anode plane, the field and grid wires will also collect induced signals from positive ions moving away from the anode wires. These ancillary electrodes  ``integrate'' signals from neighboring anode wires plus rapidly decreasing contributions from next-to-nearest neighbors, etc. If none of the secondary positive ions were collected on these electrodes, all ions would pass to the cathode plane, and signals on these electrodes would be exactly bipolar, with zero net induced image charge.  The time-integrated charge on the anode wire is indifferent to spatial variations in the arriving primary charge. 

However, because of greatly increased diffusion of electrons during approach to the anode wire, the avalanche tends to spread azimuthally around the anode wire, not shown in the field lines in Figure~\ref{fig:AnodePlane}. This spreading leads to a significant and varying fraction of ions that do not return to the cathode plane and instead land on these electrodes.  Although the signals on the ancillary electrodes are not exactly bipolar, the full-time integral of the anode-wire signal is unaffected. The ion destination can change the waveform, so finite shaping, feedback decay, or a truncated integration window can still produce an apparent amplitude variation.

Many of the well-known factors that can degrade performance are discussed in publications and documents of the Gotthard TPC experiment. Below we present an augmented list of these factors. 

\begin{enumerate}
    \item Geometrical/mechanical:
    \begin{itemize}
        \item Inaccuracies in fabrication of major parts and wires.
        \item Imprecision in assembly of major parts and wires.
        \item Inadequate tension on all wires of the proportional gain cell.

    \end{itemize}

    \item Anode wires (over long and short distances): 
    \begin{itemize}
        \item Lack of cleanliness.
        \item Non-uniformities of diameter.
        \item Non-uniformities of roundness.
        \item Inadequate smoothness.
        \item Non-optimum wire diameter for desired gain at chosen pressure.
        \item Uncompensated impact of deflections of grid wires due to differences in electric fields between drift and gain regions.
         \item Electrostatic instabilities such as even-odd displacements.

    \end{itemize}

    \item Electric fields:
    \begin{itemize}
        \item Instability/noise from power supplies.
        \item Non-circularity of equipotentials near the wire surface.
    \end{itemize}

    \item Gas system:
    \begin{itemize}
        \item Pressure instability.
        \item Anode plane geometric disturbances due to gas flow patterns.
        \item Electronegative contaminants causing attachments during drift.
        \item Presence of attachment processes at energies relevant during avalanche.
    \end{itemize}

    \item Inadequate/unstable inter-channel avalanche/electronic gain calibration among anode wires.
    \item Inadequate capture of 1/time anode signal current by electronics and post-processing software.
    \item Inadequate capture of signal components induced on ancillary electrodes in multi-wire anode designs.
    \item Electronic noise, baseline shifts, crosstalk, time constants, non-linearities, etc.
    
\end{enumerate}

While perhaps daunting in sum, each factor has limited or negligible correlations with others and may be directly addressed. Confronting these factors quantitatively must be a primary goal of future work to prepare a demonstrator phase.


\section{Machine Learning Architectures}\label{appdx:ML}

This work deploys two machine learning network architectures: (i) a Sparse 3-D Convolutional Neural Network (CNN), which is used for evaluating the $z$-position from diffusion and background rejection, and (ii) a graphical neural network (GNN) for background rejection. The code for these algorithms can be found in \cite{ATPC}. 

Common to both networks, datasets are shuffled during training, and an \texttt{AdamW} optimizer is used with weight decay. In the $z$-position reconstruction task, network optimization is done by choosing the model with the lowest validation loss. While in the background rejection task, network optimization is done by choosing the model with the largest area under the curve (AUC) metric on the validation set, which is cross-referenced with the AUC value applied on the test set to check for consistency on an unseen set of events. Class weighting is also applied to balance the signal and background datasets. 


\subsection{Sparse 3-D Convolutional Neural Network}
The CNN utilizes a voxel size of side length 8~mm. This voxel size was chosen based on the diffusion extent after drifting about 6~m. Training is done via the \texttt{spconv} library within the \texttt{PyTorch} framework. The network architecture consists of 5 sparse convolution layers (three resolution-preserving submanifold convolutions and two strided convolutions for spatial downsampling), which progressively expand the feature channels up to 128. Every convolutional layer is sequentially followed by 1-D Batch Normalization and a Rectified Linear Unit (ReLU) activation. Following the convolutions, a global pooling concatenates the per-image mean and max of the active voxel features, flattening them into a 256-dimensional vector. This representation is then fed into a 4-layer fully connected classifier (Multilayer Perceptron), which utilizes ReLU activations and a Dropout rate of 0.2 for regularization. 

In the barycentre/$z_\textnormal{max}$ position reconstruction, a mean squared error loss function is used with a numerical prediction output for the barycentre/$z_\textnormal{max}$. For the background rejection task, a binary cross entropy loss function is used to predict a signal or background label for the event. 


\subsection{Graphical Neural Network}

\begin{figure}[hbt]
\centering
\includegraphics[width=0.48\textwidth]{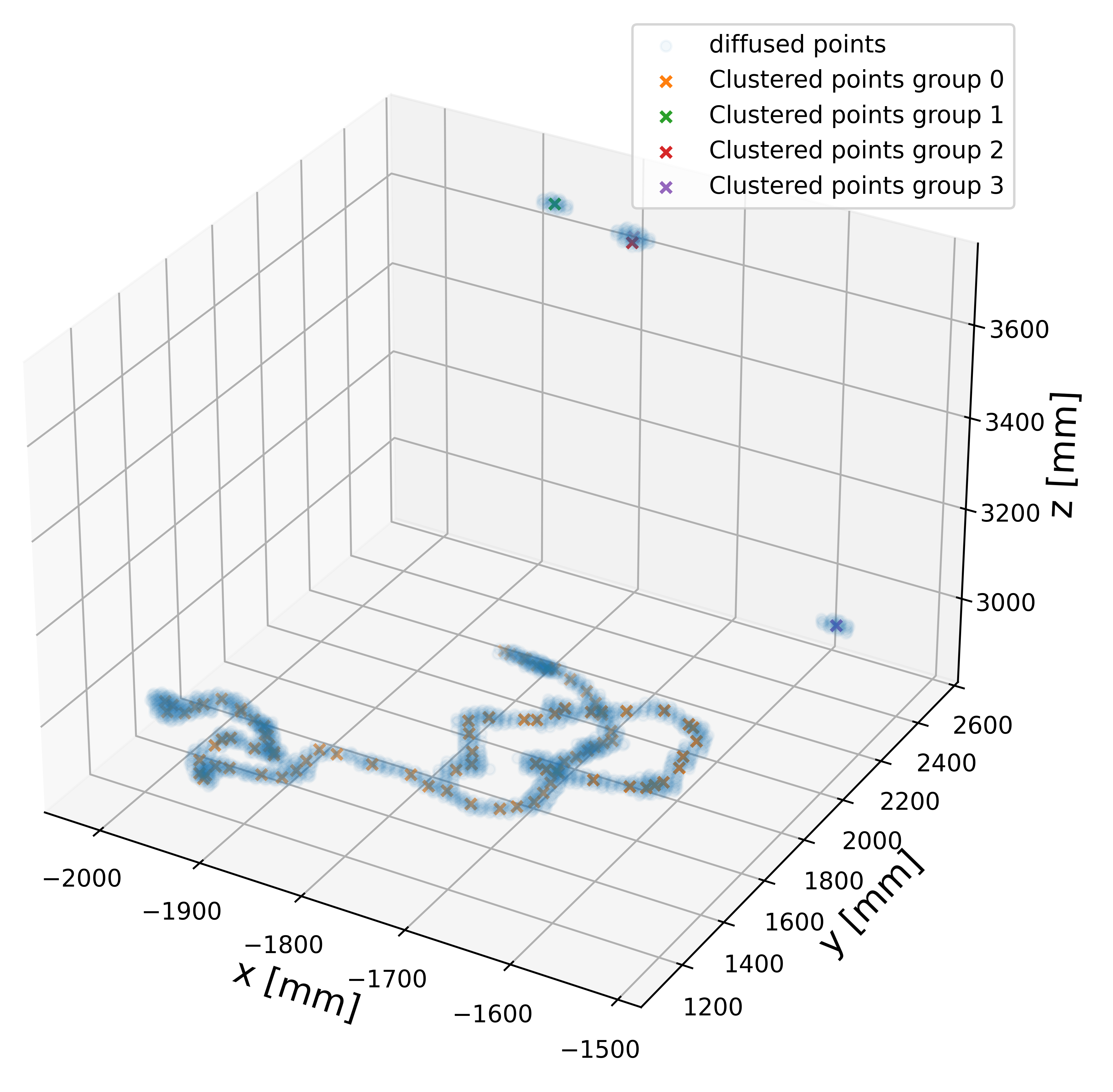}
\includegraphics[width=0.48\textwidth]{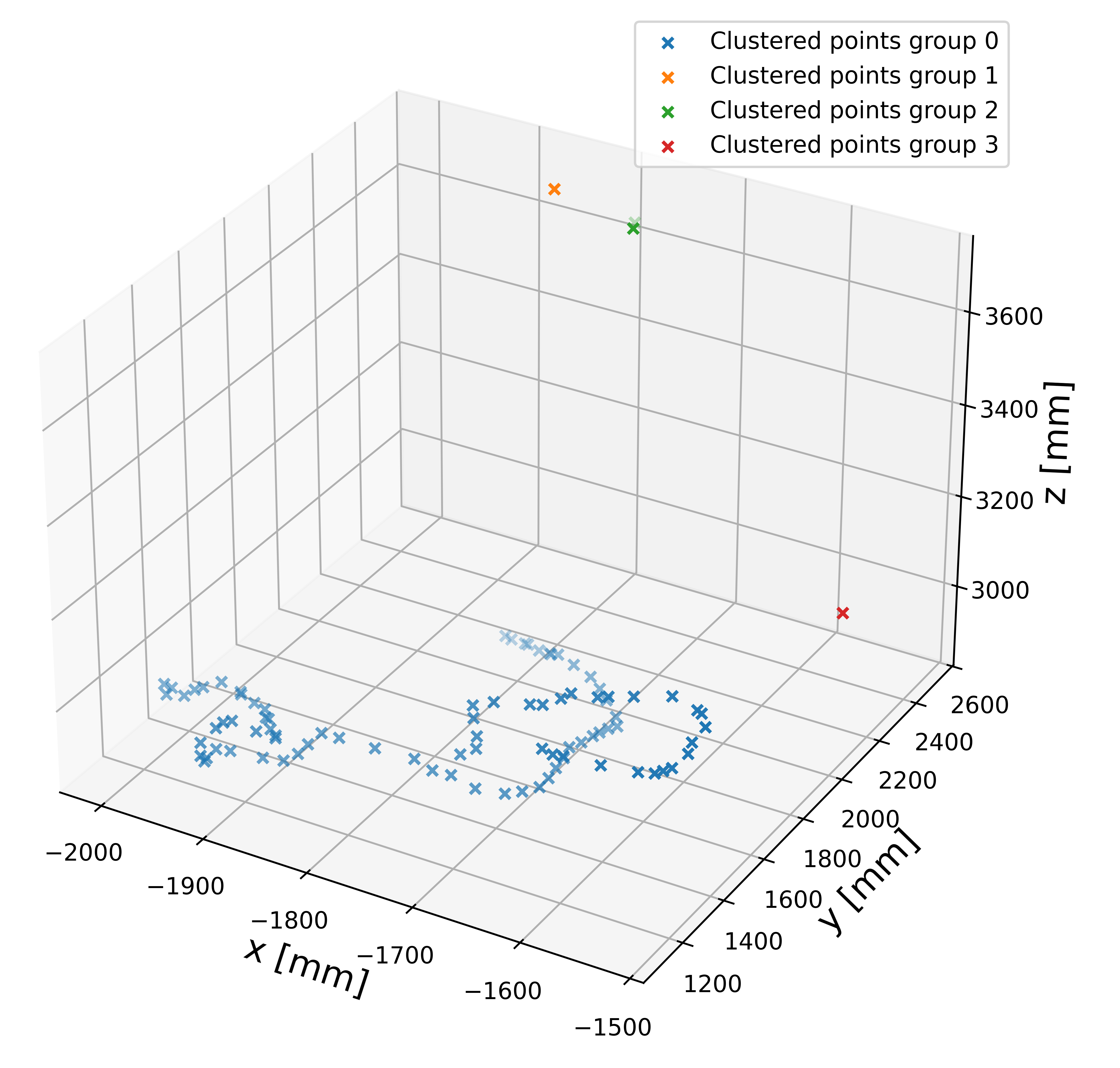}
\caption{\label{fig:clustering} An example of clustered hits for a $^{208}$Tl event. (left) shows the clustering of hits within the primary track hits in light blue. Each group is indicated by a different color cross. The right plot shows the final result of clustering without the primary track hits.     }
\end{figure}

The GNN node and edge connections are based on an algorithmic approach that derives new angular information about the tracks not previously used in gas TPC $0\nu\beta\beta$ decay searches. The diffused track is first grouped into segments based on hit locality via the \texttt{scikit-learn} \texttt{DBSCAN} function. These grouped hits are then clustered into points that find the general centre. This clustering is done by taking the median position and summing the energy of hits ordered/indexed by ($x,y,z$) lying within a sphere of a given radius determined by the expected diffusion, forming a node. This process is done until all the initial hits are accounted for. The clustering for an example $^{208}$Tl event is shown in Fig.~\ref{fig:clustering}.

Following the clustering, an algorithm then iteratively connects the nodes (for each group) to form track segments such that each node has a maximum of two connections and looping between track segments is prohibited. A second step then connects any nodes with a single connection to another node that falls within a certain distance. In this case, three connections to a given node are allowed, and again, looping between track segments is prohibited. These connections are to account for forks in a track due to $\delta$-rays. A final step walks along the track segments, labeling them:
\begin{enumerate}
    \item Primary: The track segment from end-to-end stretching the longest distance.
    \item $\delta$-ray: The track segment is attached to the primary via a node with three connections. 
    \item $\gamma$: These are track segments which are disconnected from the primary track associated with x-rays, Compton scatters, and Bremsstrahlung. These tracks can also have $\delta$-rays.
\end{enumerate}

\begin{figure}[hbt]
\centering
\includegraphics[width=\textwidth]{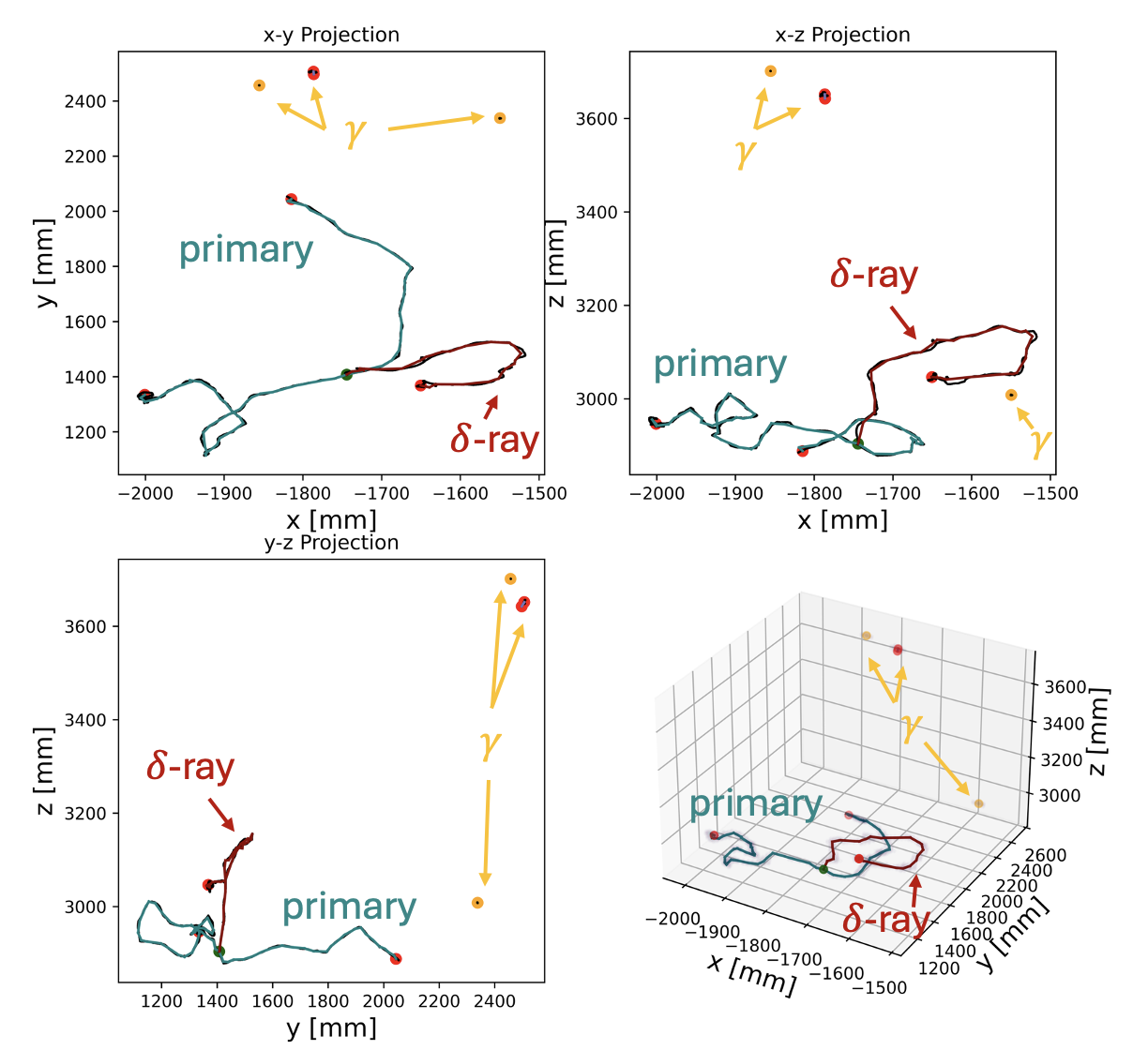}
\caption{\label{fig:trackalgoevt} An example of the tracking algorithm output following on from the same clustered hits in Fig.~\ref{fig:clustering}. The bottom right figure shows the 3-D projection, while the other panes show each 2-D projection. This event contained a high-energy $\delta$-ray indicated by the red, and several $\gamma$-ray deposits. The red circles indicate a track start/end, while the yellow circles indicate there was only one node in the deposit. In the 2-D projections, the black trace is the true (undiffused) track trajectory, while the colored lines show the reconstructed track trajectory.     }
\end{figure}

\begin{figure}[hbt]
\centering
\includegraphics[width=0.5\textwidth]{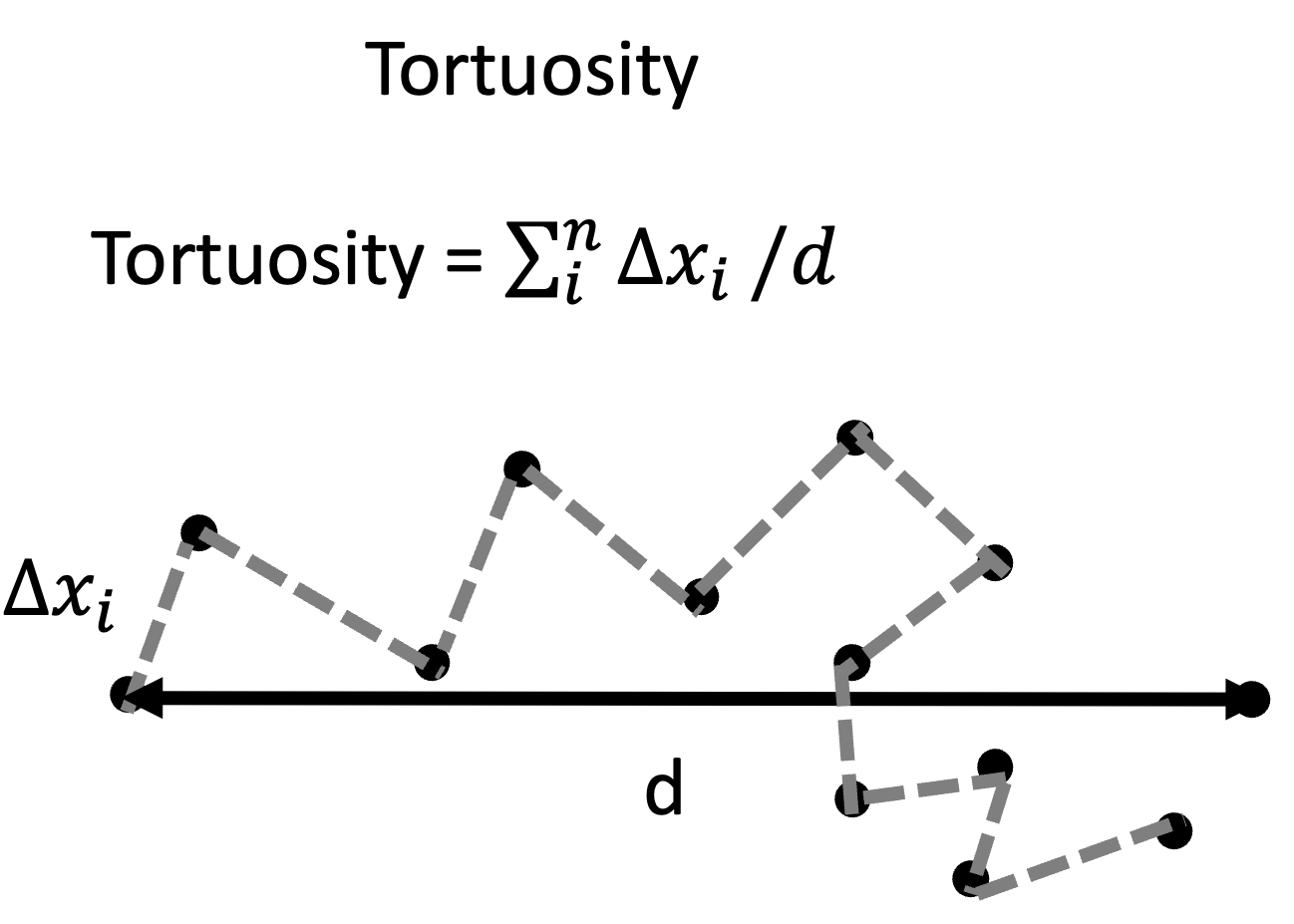}
\caption{\label{fig:tortuosity} The definition of the variable tortuosity, which characterizes how much the track segment deviates from a straight line. This variable is calculated for each node, taking a set number of nodes on either side of it in the calculation.    }
\end{figure}

Figure~\ref{fig:trackalgoevt} shows an example $^{208}$Tl event following the full reconstruction chain. Overall, the track outline and relevant features are extracted. Since this algorithm walks along a track from end to end, information such as the angular scatter can be extracted. This information is relevant for identifying stopping electrons in the track, whereby the angle of scatter increases as it slows down. Since the raw angle of scatter information from node-to-node can be noisy, we define a new variable, ``tortuosity'', which is a measure of how much a track segment deviates from a straight line. The variable is illustrated in Figure~\ref{fig:tortuosity}.

For the features of the GNN, each node is weighted by the following features: the $x$, $y$, $z$, energy, tortuosity, angle, cumulative distance along track, and label (primary, $\delta$-ray, $\gamma$). Edges include the displacement from node to node with an index of 0, and the smallest displacement from group to group with an index of 1 (to indicate the closest distance of separated deposits). 

Training is done via a \texttt{TransformerConv} GNN within the \texttt{PyTorch-Geometric} framework. \texttt{TransformerConv} was chosen due to its rapid speed via parallel operations, and good performance overall. The network architecture consists of 6 transformer convolutional layers that incorporate edge attributes during message passing. The network projects the initial node features to a fixed hidden dimension and maintains this size across all convolutions. The first five convolutional layers are followed by ReLU activations, while the sixth passes its output directly to the pooling stage. Unlike the CNN, this model did not require Batch Normalization or Dropout for regularization. Following the graph convolutions, a Global Mean Pooling operation aggregates the node features into a single, graph-level representation per event, which is then passed through a single fully connected linear layer to output the final class predictions. A binary cross entropy loss function is used to predict a signal or background label for the event.

\end{document}